\documentclass[12pt,dvips]{article}
\usepackage{rotating}
\usepackage{epsfig}
\usepackage{color}
\usepackage{cite}

\newcommand{\wt}{\widetilde}
\newcommand{\imag}{\Im {\rm m}}
\newcommand{\real}{\Re {\rm e}}

\newcommand{\lsim}{\raisebox{-0.13cm}{~\shortstack{$<$ \\[-0.07cm] $\sim$}}~}
\newcommand{\gsim}{\raisebox{-0.13cm}{~\shortstack{$>$ \\[-0.07cm] $\sim$}}~}

\begin{document}

\def\thefootnote{\fnsymbol{footnote}}

\begin{flushright}
March 2011
\end{flushright}

\begin{center}
{\bf {\Large
Higgs Mediated EDMs in the 
Next-to-MSSM : \\[3mm]
An Application to Electroweak Baryogenesis
} }
\end{center}

\medskip

\begin{center}{\large
Kingman Cheung$^{a,b,c}$,
Tie-Jiun Hou$^{a,b}$,
Jae~Sik~Lee$^b$, and
Eibun Senaha$^b$ }
\end{center}

\begin{center}
{\em $^a$ Department of Physics, National Tsing Hua University, Hsinchu, Taiwan
300}\\[0.2cm]
{\em $^b$Physics Division, National Center for Theoretical Sciences,
Hsinchu, Taiwan 300}\\[0.2cm]
{\em $^c$Division of Quantum Phases \& Devices,
                  Konkuk University,  Seoul 143-701, Korea}\\[0.2cm]
\end{center}

\bigskip\bigskip

\centerline{\bf ABSTRACT}

\noindent
We perform a study on the predictions of electric-dipole
moments (EDMs) of neutron, Mercury (Hg), Thallium (Tl), deuteron,
and Radium (Ra) in the framework of next-to-minimal supersymmetric
standard model (NMSSM) with CP-violating parameters in the superpotential
and soft-supersymmetry-breaking sector. We confine to the case in which
only the physical tree-level CP phase $(\phi'_\lambda - \phi'_\kappa)$,
associated with the couplings of the singlet terms in the superpotential
and with the vacuum-expectation-values (VEVs), 
takes on a nonzero value. We found that the one-loop contributions from
neutralinos are mostly small while the two-loop
Higgs-mediated contributions of the Barr-Zee (BZ) type diagrams dominate.
We emphasize a scenario motivated by electroweak baryogenesis.


\newpage

\section{Introduction}

Supersymmetry (SUSY) is the leading candidate for the physics beyond the
standard model (SM).  It not only solves the gauge hierarchy problem,
but also provides a dynamical mechanism for electroweak symmetry
breaking and cosmological connections such as
 a natural candidate for the dark matter and baryogenesis.
The minimal supersymmetric extension of the SM (MSSM)
has attracted much phenomenological and theoretical interests but
it suffers from the so-called
little hierarchy problem and the $\mu$ problem.

An extension with an extra singlet superfield, known as the
next-to-minimal supersymmetric standard model
(NMSSM)~\cite{NMSSM:0,NMSSM:1,NMSSM:ECPV,Boz:2005sf,Degrassi:2009yq,NMSSM:review}
was motivated to
provide a natural solution to the $\mu$ problem.  The $\mu$ parameter
in the term $\mu H_u H_d$ of the superpotential of the MSSM naturally
has its value at either $M_{\rm Planck}$ or zero (due to a symmetry).
However, the radiative electroweak symmetry breaking conditions
require the $\mu$ parameter to be of the same order as the $Z$-boson mass
for fine-tuning reasons. Such a conflict was coined as
the $\mu$ problem~\cite{mu-problem}.
In the NMSSM, the $\mu$ term is generated dynamically
through the vacuum-expectation-value (VEV), $v_S$, of the scalar
component of the additional gauge singlet Higgs superfield $\widehat{S}$, 
which is naturally of the
order of the SUSY breaking scale.  Thus, an effective $\mu$ parameter
of the order of the electroweak scale is generated.
The NMSSM was recently revived because it was shown that it can
effectively relieve the little hierarchy problem \cite{derm}.  Due to
the additional Higgs singlet field and an approximate PQ symmetry, the
NMSSM naturally has a light pseudoscalar Higgs boson $a_1$.  It has been
shown \cite{derm} that, in most parameter space that is natural, the
SM-like Higgs boson can decay into a pair of light pseudoscalar bosons
with a branching ratio larger than $0.7$.  Thus, the branching ratio
of the SM-like Higgs boson into $b\bar b$ would be less than $0.3$ and
so the LEPII bound is effectively reduced to around 100 GeV
\cite{LEP2003}.  Since the major decay modes of the Higgs boson are
no longer $b\bar b$, unusual search modes have been investigated
\cite{nmssm-new}.

CP violation is one of the necessary ingredients 
for successful baryogenesis~\cite{Sakharov:1967dj}.
Although the SM can accommodate CP violation originating from the 
Cabibbo-Kobayashi-Maskawa matrix~\cite{CKM},  
it turns out that its effect is way too small to generate 
sufficient baryon asymmetry ($\sim10^{-10}$)~\cite{ewbg_sm_cp}
\footnote{Another shortcoming in the SM baryogenesis is that 
the electroweak phase transition is a smooth crossover for $m_h>$ 
73 GeV~\cite{sm_ewpt}, 
rendering thermal nonequilibrium unrealizable.}.
This fact suggests, in turn, that there should be an extra source of CP violation 
which has not been probed yet. 
CP violation relevant to electroweak baryogenesis (EWBG)~\cite{ewbg} 
by construction must appear 
in the Higgs self interactions and/or the Higgs interactions with the 
other particles whose masses are $ \mathcal{O}(100)$ GeV. 
Therefore, such CP violating effects can be communicated to the 
low energy observables
which are measurable in the near future experiments. 
A lot of effort on the EWBG study have been made in the new physics models 
such as the MSSM~\cite{ewbg-mssm}, the two-Higgs doublet model~\cite{ewbg-2hdm}
and the singlet-extended MSSM~\cite{Funakubo:2005pu,ewbg-xMSSM}.

The MSSM offers many possible sources of CP violation beyond the 
single Kobayashi-Maskawa phase in the SM.
As far as the Higgs sector is concerned,
the non-vanishing CP phases could induce
significant mixing between the CP-even and
CP-odd states radiatively~\cite{CPmixing0,CPmixing1,CPmixing1.5,CPmixing2},
giving rise to a number of interesting CP violating phenomena
and substantial modifications to
Higgs-boson phenomenology~\cite{Lee:2008eqa,Accomando:2006ga}.
In particular, the lightest Higgs boson can be as light as a few GeV
with almost vanishing couplings to the weak gauge bosons
when the CP-violating phases are maximal.  The decay patterns of the
heavier Higgs bosons become much more
complicated compared to the CP-conserving case
because of the loss of its CP parities~\cite{CPH_decay,cpsuperh}.
These combined features make the Higgs boson searches at LEP difficult,
consequently, the Higgs boson lighter than $\sim$ 50 GeV
can survive the LEP limit~\cite{Schael:2006cr}.

The non-observation of electric dipole moments (EDMs) for
Thallium~\cite{Regan:2002ta}, neutron~\cite{Baker:2006ts},
and Mercury~\cite{Romalis:2000mg,Griffith:2009zz} is known to
constrain the CP-violating phases very tightly. 
It is generally believed that one-loop contributions
dominate and we set the phases to ${\cal O}(0)$ 
to make the null results of the EDM searches consistent
within most of the parameter space
\footnote{Nevertheless,
accidental cancellations among various contributions may occur
in the three measured EDMs,
thus still allowing sizable CP phases
even with the SUSY particles lighter than
${\cal O}(1~{\rm TeV})$~\cite{Ibrahim:1998je,Ellis:2008zy}.}.  
However, we point out in this work
that it may not be the case in the framework of NMSSM with CP-violating
parameters. Even if we set the CP phases of the parameters
appearing in the MSSM to zero, there could be 
potentially large nontrivial two-loop contributions 
coming from a combination of the CP phases,
$(\phi'_\lambda - \phi'_\kappa)$,
which could exist only in the NMSSM.
We note that, being different from the MSSM, the non-vanishing
CP phase could cause CP-violating mixing among the neutral 
Higgs bosons even at the tree level.

In this work, we perform a study on the predictions
of EDMs of neutron, Mercury (Hg), Thallium (Tl), deuteron, and Radium (Ra)
in the framework of NMSSM with CP-violation.
We confine ourselves to the case in which
only the physical CP phase $(\phi'_\lambda - \phi'_\kappa)$,
associated with the couplings of the terms containing the singlet field
in the superpotential and with the VEVs, takes on a nonzero value.
We figure out how large the CP phase can be
taken in a scenario in which
a first-order phase transition could be achieved more easily
in comparison to the MSSM~\cite{Funakubo:2005pu}.
The form factors that contribute to these observable EDMs include
electric-dipole moment, chromo-electric dipole moment, Weinberg
three-gluon operator, and the four-fermion operators.
The two-loop Weinberg three-gluon operator
and the Higgs-exchange four-fermion operators are generated
due to the tree-level CP-violating Higgs mixing.
The electric-dipole moment (EDM) and
chromo-electric-dipole moment (CEDM) receive 
the following one- and two-loop contributions:
(i) One loop neutralino-sfermion contribution in which the CP phase appears
in the neutralino mass matrix (the CP phase of the effective $\mu$ parameter
in the chargino-mass matrix is set to zero).
(ii) Two-loop Barr-Zee (BZ) diagram with the
$\gamma H^0$, $W^\mp H^\pm$, $W^\mp W^\pm$, and $Z H^0$ decompositions.

We found that the one-loop contributions from
the neutralino-sfermion diagrams to the Thallium and neutron EDMs
lie below the present experimental upper limits
especially when the sfermions of the first two generations
are heavier than $\sim$ 300 GeV.  This is in agreement with the 
previous observations~\cite{Funakubo:2004ka,Boz:2005sf}, in which only the 
one-loop contributions were taken into account.
The one-loop contributions to the Mercury EDM could be larger 
but they also go below the present experimental upper limit
if the sfermions of the first two generations
are heavier than $\sim 300 - 500$ GeV.

The two-loop contributions start to dominate
when the sfermions of the first two generations
are heavier than $\sim$ 300 GeV and
the one-loop contributions are suppressed.
We found that the two-loop contributions
can saturate the current bound on the neutron EDM
and they can go over that on the Mercury EDM.
But we found that there is still a room to have the maximal
CP phase $(\phi_\lambda^\prime-\phi_\kappa^\prime) \sim 90^\circ$
after taking account of the uncertainties in the calculations of
the EDMs.
We note that the large CP phase can be easily probed in
the proposed future experiments searching for the EDMs of
the deuteron and the $^{225}$Ra atom and it might be
connected to the EWBG.
%

The organization of the paper
is as follows. We briefly describe the Higgs sector
of the NMSSM with CP-violating  parameters in Sec. II.  We give the
relevant Higgs couplings in Sec. III and detail breakdowns of the EDM
calculations in Sec. IV. Numerical analysis is given in Sec. V. We
conclude in Sec. VI.

\section{Higgs sector in the NMSSM with CP violation}
The superpotential of the NMSSM may be written as
\begin{equation}
  \label{Wpot}
W_{\rm NMSSM}\ =\ \widehat{U}^C {\bf h}_u \widehat{Q} \widehat{H}_u\:
+\:   \widehat{D}^C {\bf h}_d \widehat{H}_d \widehat{Q}  \: +\:
\widehat{E}^C {\bf h}_e \widehat{H}_d \widehat{L} \: +\:
\lambda \widehat{S} \widehat{H}_u \widehat{H}_d\ \: + \:
\frac{\kappa}{3}\ \widehat{S}^3 \ ,
\end{equation}
where $\widehat{S}$ denotes the singlet Higgs superfield,
$\widehat{H}_{u,d}$  are the two SU(2)$_L$ doublet Higgs superfields, and
$\widehat{Q}$,  $\widehat{L}$ and $\widehat{U}^C$,  $\widehat{D}^C$,
$\widehat{E}^C$ are the matter doublet and singlet superfields, respectively,
related to up- and  down-type quarks and  charged leptons.
We note that, especially, the last cubic term
with a dimensionless coupling $\kappa$
respects an extra discrete $Z_3$ symmetry.
The superpotential leads to the tree-level Higgs potential, which
is given by the sum
\begin{eqnarray}
V_0=V_F+V_D+V_{\rm soft},
\end{eqnarray}
where each term is given by
\begin{eqnarray}
V_F&=&|\lambda|^2|S|^2(H_d^\dagger H_d+H_u^\dagger H_u)
        +|\lambda H_u H_d+\kappa S^2|^2,\nonumber \\
V_D&=&\frac{g^{\prime 2}+g^2}{8}(H_d^\dagger H_d-H_u^\dagger H_u)^2
        +\frac{g^2}{2}(H_d^\dagger H_u)(H_u^\dagger H_d),\nonumber \\
V_{\rm soft}&=&m_1^2 H_d^\dagger H_d+m_2^2 H_u^\dagger H_u
        +m_S^2|S|^2
        +\left(\lambda A_{\lambda}S H_u H_d
        -\frac{1}{3}\kappa A_\kappa S^3+{\rm h.c.}\right),
\end{eqnarray}
with the gauge-coupling constants 
$g^\prime=e/\cos\theta_W$ and $g=e/\sin\theta_W$,
where $e$ is the electric charge of the positron in our convention.
Note that we are taking
the unusual minus($-$) sign for the singlet soft-trilinear term proportional
to $A_\kappa$.
%

We have parametrized the component fields of
the two doublet and one singlet scalar Higgs fields and
the vacuum expectation values (VEVs) as follows,
\begin{eqnarray}
H_d&=&
\hphantom{e^{i\theta}}\,
\left(
\begin{array}{c}
\frac{1}{\sqrt{2}}\,(v_d+\phi^0_d+ia_d) \\
\phi_d^-
\end{array}
\right), \nonumber \\[1mm]
H_u&=&
e^{i\theta}\,\left(
\begin{array}{c}
\phi_u^+\\
\frac{1}{\sqrt{2}}\,(v_u+\phi^0_u+ia_u)
\end{array}
\right), \nonumber \\[1mm]
S&=&\frac{e^{i\varphi}}{\sqrt{2}}\,(v_S+\phi^0_S+ia_S)\,.
\label{eq:higgsparam}
\end{eqnarray}
Note that we have the complex vacuum-expectation-values (VEVs) and assume
that the parameters $\lambda$ and $\kappa$ in the superpotential and
$A_\lambda$ and $A_\kappa$ in the soft terms contain non-trivial CP phases.
It turns out that
not all the CP phases appearing at the tree level 
after the electroweak symmetry breaking 
are physical and the only physical one is the difference 
$\phi^\prime_\lambda - \phi^\prime_\kappa$ with
\begin{equation}
\phi^\prime_\lambda \equiv \phi_\lambda+\theta+\varphi \ \ \ {\rm and} \ \ \
\phi^\prime_\kappa \equiv \phi_\kappa+3\varphi\,,
\end{equation}
and the CP phases of $A_\lambda$ and $A_\kappa$ are determined 
up to a two-fold ambiguity using the two CP-odd tadpole conditions.
When $\phi^\prime_\lambda - \phi^\prime_\kappa \neq 0$, 
the neutral Higgs bosons do not have to carry any definite CP parities
already at the tree level and its mixing is described by
the orthogonal $5\times 5$ matrix $O_{\alpha i}$ as
\begin{equation}
\left( \phi^0_d\,, \phi^0_u\,, \phi^0_S\,, a\,, a_S \right)^T \ = \
O_{\alpha i}
\left( H_1\,, H_2\,, H_3\,, H_4\,, H_5 \right)^T
\end{equation}
with $H_{1(5)}$  the lightest (heaviest) Higgs mass eigenstate.

For the calculation of the Higgs-boson masses and mixing matrix $O_{\alpha i}$
in the presence of CP-violating parameters in the superpotential and in the
soft-supersymmetry-breaking sector,
we adopt the renormalization-group (RG) improved approach
by including the full one-loop
and the logarithmically enhanced two-loop effects~\cite{Cheung:2010ba}.
%
And then, the NMSSM Higgs sector is fixed by specifying the following
input parameters:
\begin{eqnarray}
{\rm tree~level} &:&  |\lambda|\,,|\kappa|\,,\tan\beta\,; \
|A_\lambda|\,,|A_\kappa|\,,v_S
\nonumber \\
\mbox{ 1-loop~level} &:&
M_{{\widetilde Q}_3}\,,M_{{\widetilde U}_3}\,,M_{{\widetilde D}_3}\,,
|A_t|\,,|A_b|
\nonumber \\
{\rm CP~phases} &:& \phi_\lambda^\prime\,,\phi_\kappa^\prime\ ; \ \phi_{A_t}\,,\phi_{A_b}
\nonumber \\
{\rm signs~of} &:& \cos(\phi_\lambda^\prime+\phi_{A_\lambda})\,,
\cos(\phi_\kappa^\prime+\phi_{A_\kappa})\,.
\end{eqnarray}
For the renormalization scale $Q_0$ we take the top-quark mass
as in Refs.~\cite{Casas:1994us,CPmixing1,CPmixing2}.

In this work, we wish to consider the constraint on the tree-level
CP phase $\phi^\prime_\lambda - \phi^\prime_\kappa$ coming from
the non-observation  of  electric dipole  moments (EDMs)  for Thallium
($^{205}{\rm Tl}$)~\cite{Regan:2002ta}, the neutron~($n$)~\cite{Baker:2006ts},     
and Mercury ($^{199}{\rm Hg}$)~\cite{Romalis:2000mg,Griffith:2009zz} when the CP phases
appearing in all the other soft SUSY-breaking terms vanish or
$\sin(\phi^\prime_\lambda+\phi_{A_f})=\sin(\phi^\prime_\lambda+\phi_i)=0$,
with $\phi_{A_f}$ and $\phi_i$ denoting the CP phases of the soft 
trilinear parameters $A_f$ and the three gaugino mass 
parameters $M_{i=1,2,3}$, respectively.

\section{Higgs-boson couplings in the NMSSM}

If we consider the case in which only the tree-level
CP phase $\phi^\prime_\lambda - \phi^\prime_\kappa$ 
takes a non-trivial value while all the other CP phases 
are vanishing, as will be shown in the 
following, the one-loop contributions to the EDMs 
are mostly small and the EDMs are dominated by the 
two-loop contributions from the Higgs-mediated  dimension-6 Weinberg
operator,  the Higgs-exchange four-fermion  operators, and
the Barr--Zee-type diagrams.
Therefore, for the  calculation of the EDMs 
beyond the one-loop level,
one may need the Higgs-boson couplings taking full account of the 
$5 \times 5$ CP-violating mixing matrix $O_{\alpha i}$.
In this section,
we present the couplings of the neutral and charged Higgs bosons to quarks, 
leptons, charginos, neutralinos, and third-generation
sfermions in the NMSSM with CP violation.
For the conventions and notations of the masses and mixing
matrices of charginos, neutralinos, and third-generation
sfermions, we refer to Appendix A.
%

The interactions of the five neutral Higgs bosons with the SM quarks and leptons
are described by the interaction Lagrangian:
\begin{equation}
{\cal L}_{H_i\bar{f}f}\ =\ - \,g_f\,
\sum_{i=1}^5\, H_i\, \bar{f}\,\Big( g^S_{H_i\bar{f}f}\, +\,
ig^P_{H_i\bar{f}f}\gamma_5 \Big)\, f\ ,
\end{equation}
where $g_f=g m_f/2 M_W$ for $f=u,d,l$.
At the tree level, $(g^S_{H_i\bar{f}f},g^P_{H_i\bar{f}f}) = (O_{1 i}/c_\beta, -O_{4i}
\tan\beta)$ and 
$(g^S_{H_i\bar{f}f}, g^P_{H_i\bar{f}f}) = (O_{2 i}/s_\beta, -O_{4i}
\cot\beta)$ for $f=(l,d)$ and $f=u$, respectively.
The simultaneous existence of the couplings $g^S_{H_i\bar{f}f}$ and
$g^P_{H_i\bar{f}f}$ signals CP violation.
The couplings of the neutral Higgs bosons to the five neutralinos are given by:
\begin{eqnarray}
{\cal L}_{H^0\wt{\chi}^0\wt{\chi}^0}
&=&-\frac{g}{2}\sum_{i,j,k} H_k
\overline{\wt{\chi}_i^0}
\left(g_{H_k\widetilde{\chi}^0_i\widetilde{\chi}^0_j}^{S}
+i\gamma_5
g_{H_k\widetilde{\chi}^0_i\widetilde{\chi}^0_j}^{P}\right)
\wt{\chi}_j^0 \,
\end{eqnarray}
where $i,j=1,2,3,4,5$ for the five neutralinos and $k=1,2,3,4,5$ for the five
neutral Higgs bosons and the scalar and pseudo-scalar coupling are
\begin{eqnarray}
g_{H_k\widetilde{\chi}^0_i\widetilde{\chi}^0_j}^{S} &=&\real
\Bigg\{
\frac{1}{2}(N_{j2}^*-t_W N_{j1}^*)
[N_{i3}^*(O_{1k}-iO_{4k}s_\beta)-
N_{i4}^*(O_{2k}-iO_{4k}c_\beta)]
\nonumber \\ && \hspace{0.5cm}
-\frac{|\lambda| e^{i(\phi_\lambda+\theta+\varphi)}}{\sqrt{2}\,g}
\left[
(O_{3k}+iO_{5k})N_{i4}^*N_{j3}^*
+(O_{2k}+iO_{4k}c_\beta )N_{i5}^*N_{j3}^* \right.
\nonumber \\ && \hspace{3.2cm} \left.
+(O_{1k}+iO_{4k}s_\beta )N_{i5}^*N_{j4}^* \right]
\nonumber \\ && \hspace{0.5cm}
+\frac{|\kappa| e^{i(\phi_\kappa+3\varphi)}}{\sqrt{2}\,g}
\left[ (O_{3k}+iO_{5k})N_{i5}^*N_{j5}^* \right]
+[i\leftrightarrow j]\Bigg\}
\nonumber \\
g_{H_k\widetilde{\chi}^0_i\widetilde{\chi}^0_j}^{P} &=&-\imag
\Bigg\{
\frac{1}{2}(N_{j2}^*-t_W N_{j1}^*)
[N_{i3}^*(O_{1k}-iO_{4k}s_\beta)-
N_{i4}^*(O_{2k}-iO_{4k}c_\beta)]
\nonumber \\ && \hspace{0.5cm}
-\frac{|\lambda| e^{i(\phi_\lambda+\theta+\varphi)}}{\sqrt{2}\,g}
\left[
(O_{3k}+iO_{5k})N_{i4}^*N_{j3}^*
+(O_{2k}+iO_{4k}c_\beta )N_{i5}^*N_{j3}^* \right.
\nonumber \\ && \hspace{3.2cm} \left.
+(O_{1k}+iO_{4k}s_\beta )N_{i5}^*N_{j4}^* \right]
\nonumber \\ && \hspace{0.5cm}
+\frac{|\kappa| e^{i(\phi_\kappa+3\varphi)}}{\sqrt{2}\,g}
\left[ (O_{3k}+iO_{5k})N_{i5}^*N_{j5}^* \right]
+[i\leftrightarrow j]\Bigg\}\,.
\end{eqnarray}
Note the couplings are symmetric under the exchange of 
$i\leftrightarrow j$, reflecting the
Majorana property of the neutralinos, and contain the terms
coupled to the singlet components of the Higgs bosons and 
to the singlino components of the neutralinos which do not exist
in the MSSM.
The couplings of the neutral Higgs bosons to the charginos
can be similarly cast into the form:
\begin{eqnarray}
{\cal L}_{H^0\wt{\chi}^+\wt{\chi}^-}
&=&-\frac{g}{\sqrt{2}}\sum_{i,j,k} H_k
\overline{\wt{\chi}_i^-}
\left(g_{H_k\widetilde{\chi}^+_i\widetilde{\chi}^-_j}^{S}+i\gamma_5
g_{H_k\widetilde{\chi}^+_i\widetilde{\chi}^-_j}^{P}\right)
\wt{\chi}_j^-\,,
\end{eqnarray}
with $i,j=1,2$ for the two charginos, $k=1,2,3,4,5$ for the five
neutral Higgses, and
\begin{eqnarray}
g_{H_k\widetilde{\chi}^+_i\widetilde{\chi}^-_j}^{S}&=&\frac{1}{2}
\Bigg\{
(C_R)_{i1}(C_L)^*_{j2}(O_{1k}-iO_{4k}s_\beta)+
(C_R)_{i2}(C_L)^*_{j1}(O_{2k}-iO_{4k}c_\beta)
\nonumber \\ && \hspace{0.5cm}
+\frac{|\lambda| e^{i(\phi_\lambda+\theta+\varphi)}}{g}
(C_R)_{i2}(C_L)_{j2}^*(O_{3k}+iO_{5k})
+[i\leftrightarrow j]^* \Bigg\}\,,
\nonumber \\
g_{H_k\widetilde{\chi}^+_i\widetilde{\chi}^-_j}^{P} &=&\frac{i}{2}
\Bigg\{
(C_R)_{i1}(C_L)^*_{j2}(O_{1k}-iO_{4k}s_\beta)+
(C_R)_{i2}(C_L)^*_{j1}(O_{2k}-iO_{4k}c_\beta)
\nonumber \\ && \hspace{0.5cm}
+\frac{|\lambda| e^{i(\phi_\lambda+\theta+\varphi)}}{g}
(C_R)_{i2}(C_L)_{j2}^*(O_{3k}+iO_{5k})
-[i\leftrightarrow j]^* \Bigg\}\,.
\end{eqnarray}
We observe the couplings are real when $i=j$ but they are complex
otherwise with 
$g_{H_k\widetilde{\chi}^+_2\widetilde{\chi}^-_1}^{S,P}=
\left(g_{H_k\widetilde{\chi}^+_1\widetilde{\chi}^-_2}^{S,P}\right)^*$.
We again note that the couplings contain the terms coupled to
the singlet components of the neutral Higgs bosons which do not
exist in the MSSM.
Lastly, the neutral Higgs interactions with sfermions can be written in terms 
of the sfermion mass eigenstates as
\begin{equation}
{\cal L}_{H\widetilde{f}\widetilde{f}}= v \sum_{f=u,d,l,\nu}\,
g_{H_i \widetilde{f}^*_j \widetilde{f}_k}\,
\left(H_i\, \widetilde{f}^*_j\, \widetilde{f}_k\right)\,
\end{equation}
where
\begin{equation}
v \, g_{H_i \widetilde{f}^*_j \widetilde{f}_k} =
\left(\Gamma^{\alpha \widetilde{f}^*\widetilde{f}}\right)_{\beta\gamma}\,
O_{\alpha i} (U^{\widetilde{f}}_{\beta j})^* U^{\widetilde{f}}_{\gamma k}
\end{equation}
with $ \alpha = (\phi_d^0,\phi_u^0,\phi_S^0,a,a_S)=(1,2,3,4,5) \,,
\beta\,,\gamma = (\widetilde{f}_L, \widetilde{f}_R)=(1,2) \,,
i=(H_1,H_2,H_3,H_4,H_5)=(1,2,3,4,5) \,, {\rm and} \
j\,,k=(\widetilde{f}_1,\widetilde{f}_2)=(1,2) $.
The explicit expressions of the couplings 
$\Gamma^{\alpha \widetilde{f}^*\widetilde{f}}$ in the weak basis are given
in Appendix B.

Now let us move to the couplings of the charged Higgs bosons.
The charged Higgs boson couplings to the SM quarks and leptons are
described by the Lagrangian
\begin{equation} \label{Hud}
{\cal L}_{H^\pm f_{\uparrow}f_{\downarrow}}\ =\ -\,
g_{f_\uparrow f_\downarrow}\,
H^+\, \bar{f}_{\uparrow}\,\Big(\,
g^S_{H^+\bar{f}_{\uparrow}f_{\downarrow}}\ + i \
g^P_{H^+\bar{f}_{\uparrow}f_{\downarrow}}\ \gamma_5
\Big)\, f_{\downarrow}\ +\ {\rm h.c.} ,
\end{equation}
with $g_{f_\uparrow f_\downarrow}=-gm_u/\sqrt{2}m_W$ and
$-gm_l/\sqrt{2}m_W$ when
$(f_{\uparrow},f_{\downarrow})=(u,d)$ and $(\nu,l)$, respectively.
At the tree level,
$(g^S_{H^+\bar{u}d}\,,g^P_{H^+\bar{u}d})
=([1/t_\beta+(m_d/m_u)t_\beta)]/2\,,i[1/t_\beta-(m_d/m_u)t_\beta)]/2)$ and
$(g^S_{H^+\bar{\nu}l}\,,g^P_{H^+\bar{\nu}l})
=(t_\beta/2\,,-it_\beta/2)$.
The interactions of the charged Higgs bosons with charginos and neutralinos
are described by the following Lagrangian:
\begin{eqnarray}
{\cal L}_{H^\pm\wt{\chi}_i^0\wt{\chi}_j^\mp}
&=&-\frac{g}{\sqrt{2}}\sum_{i,j} H^+
\,\overline{\wt{\chi}_i^0}
\left(g_{H^+\widetilde{\chi}^0_i\widetilde{\chi}^-_j}^{S}
+i\gamma_5
g_{H^+\widetilde{\chi}^0_i\widetilde{\chi}^-_j}^{P}\right)
\wt{\chi}_j^-
+{\rm h.c.}\,,
\end{eqnarray}
with $i=1,2,3,4,5$, $j=1,2$, and
\begin{eqnarray}
g_{H^+\widetilde{\chi}^0_i\widetilde{\chi}^-_j}^{S}
&=&\frac{1}{2}
\left\{s_\beta\left[\sqrt{2}N_{i3}^*(C_L)_{j1}^*-(N_{i2}^*
+t_W N_{i1}^*)(C_L)_{j2}^*\right] \right.
\nonumber \\ &&~
+\frac{\sqrt{2}|\lambda| e^{i(\phi_\lambda+\theta+\varphi)}}{g}
c_\beta N_{i5}^*(C_L)_{j2}^* \nonumber \\
&&~  +c_\beta\left[\sqrt{2}N_{i4}(C_R)_{j1}^*+(N_{i2}
+t_W N_{i1})(C_R)_{j2}^*\right]
\nonumber \\ &&\left. ~
+\frac{\sqrt{2}|\lambda| e^{-i(\phi_\lambda+\theta+\varphi)}}{g}
s_\beta N_{i5}(C_R)_{j2}^*
\right\} \,, \nonumber \\
g_{H^+\widetilde{\chi}^0_i\widetilde{\chi}^-_j}^{P}
&=&\frac{i}{2}
\left\{s_\beta\left[\sqrt{2}N_{i3}^*(C_L)_{j1}^*-(N_{i2}^*
+t_W N_{i1}^*)(C_L)_{j2}^*\right] \right.
\nonumber \\ &&~
+\frac{\sqrt{2}|\lambda| e^{i(\phi_\lambda+\theta+\varphi)}}{g}
c_\beta N_{i5}^*(C_L)_{j2}^* \nonumber \\
&&~  -c_\beta\left[\sqrt{2}N_{i4}(C_R)_{j1}^*+(N_{i2}
+t_W N_{i1})(C_R)_{j2}^*\right]
\nonumber \\ &&\left. ~
-\frac{\sqrt{2}|\lambda| e^{-i(\phi_\lambda+\theta+\varphi)}}{g}
s_\beta N_{i5}(C_R)_{j2}^*
\right\}\,.
\end{eqnarray}
As similarly in the neutral Higgs couplings,
we note that the couplings are containing, 
in addition to the corresponding MSSM interactions, the terms coupled 
to the singlino components of the neutralinos.
Finally, the charged Higgs couplings to sfermions are given by
\begin{equation}
{\cal L}_{H^\pm\widetilde{f^\prime}\widetilde{f}}=
v \, g_{H^+ \widetilde{f}^*_j \widetilde{f^\prime}_k}\,
\left(H^+\, \widetilde{f}^*_j\, \widetilde{f^\prime}_k\right)\, +{\rm h.c.}
\end{equation}
where
\begin{equation}
v \, g_{H^+ \widetilde{f}^*_j \widetilde{f^\prime}_k} =
\left(\Gamma^{H^+ \widetilde{f}^*\widetilde{f^\prime}}\right)_{\beta\gamma}\,
(U^{\widetilde{f}}_{\beta j})^* U^{\widetilde{f^\prime}}_{\gamma k}
\end{equation}
The explicit expressions of the
couplings $\Gamma^{H^+ \widetilde{f}^*\widetilde{f}^\prime}$ in the weak basis
are also given in Appendix B.

\section{Synopsis of EDMs}
In this section, we briefly outline how we estimate
observable EDMs.
We start by giving the relevant interaction Lagrangian as follows:
\begin{eqnarray}
{\cal L} & = &
  -\; \frac{i}{2}\,d^E_f\,F^{\mu\nu}\,\bar{f}\,\sigma_{\mu\nu}\gamma_5\,f\
-\ \frac{i}{2}\,d_q^C\,G^{a\,\mu\nu}\,\bar{q}\,\sigma_{\mu\nu}\gamma_5 T^a\,q
\nonumber \\
 &&
+\frac{1}{3}\,d^{\,G}\,f_{abc}\,G^a_{\rho\mu}\,\widetilde{G}^{b\,\mu\nu}\,
{G^c}_{\nu}^{~~\rho}\ + \
\sum_{f,f'}     C_{ff'}     (\bar{f}f)
(\bar{f'}i\gamma_5f')\; ,
\end{eqnarray}
where
$F^{\mu\nu}$ and  $G^{a\,\mu\nu}$ are  the  electromagnetic and
strong  field strengths, respectively,  the $T^a=\lambda^a/2$  are the
generators    of    the     SU(3)$_C$    group and
$\widetilde{G}^{\mu\nu} = \frac{1}{2} \epsilon^{\mu\nu\lambda\sigma}
G_{\lambda\sigma}$ is the dual  of the SU(3)$_c$ field-strength tensor
$G_{\lambda\sigma}$.
The EDMs and the CEDMs of quarks and leptons are denoted by
$d^E_f$ and $d_q^C$, respectively.

For the Weinberg operator, we consider the contributions from 
the Higgs-mediated two-loop diagrams:
\begin{equation}
(d^{\,G})^{H}\ =\ \frac{4\sqrt{2}\, G_F\, g_s^3}{(4\pi)^4}
\sum_{q=t,b} \left[\sum_i g^S_{H_i\bar{q}q}\,
  g^P_{H_i\bar{q}q}\,h(z_{iq})\right]\,,
\end{equation}
where $z_{iq} \equiv M_{H_i}^2/m_q^2$ and, for the loop function $h(z_{iq})$,
we refer to Ref.~\cite{Dicus:1989va}.

For the four-fermion operators, we consider
the $t$-channel exchanges of the CP-violating neutral Higgs bosons 
which give rise to the CP-odd coefficients as follows~\cite{Ellis:2008zy}:
\begin{equation}
  \label{eq:cff}
(C_{ff'})^H\ =\ g_f\, g_{f'}\,\sum_i
\frac{g^S_{H_i\bar{f}f}\,g^P_{H_i\bar{f'}f'}}{M_{H_i}^2}\; .
\end{equation}
%

The EDM $d^E_f$ and CEDM $d_q^C$ are given by
the sum of the one-loop and two-loop contributions 
\begin{eqnarray}
d_f^E  =  (d^E_f)^{\widetilde\chi^0} + (d^E_f)^{\rm BZ}\,; \ \ \
d_q^C  =  (d^C_q)^{\widetilde\chi^0} + (d^C_q)^{\rm BZ}\,.
\end{eqnarray}
The details of the neutralino-mediated one-loop contributions
and the contributions from the two-loop
Barr--Zee-type diagrams will be discussed below.

\subsection{One-loop EDMs}
In the case under consideration, the only
non-vanishing one-loop contribution to the (C)EDMs comes from the neutralino loops
due to the CP phase $\phi^\prime_\lambda - \phi^\prime_\kappa$.
The  one-loop contributions to  the EDMs  of
charged    leptons   $(d_l^E/e)^{\widetilde{\chi}^0}$,    up-type   quarks
$(d_u^E/e)^{\widetilde{\chi}^0}$        and        down-type        quarks
$(d_d^E/e)^{\widetilde{\chi}^0}$ may conveniently be expressed as
\begin{equation}
\left(\frac{d^E_f}{e}\right)^{\widetilde\chi^0}\ =\
\frac{1}{16\pi^2}\sum_{i=1}^5\sum_{j=1}^2 \frac{m_{\widetilde{\chi}^0_i}}{m_{\widetilde{f}_j}^2}\,
\imag[(g_{R\,ij}^{\widetilde{\chi}^0 f \widetilde{f}})^*\,
g_{L\,ij}^{\widetilde{\chi}^0 f \widetilde{f}}]\,
Q_{\widetilde{f}}\, B({m_{\widetilde{\chi}^0_i}^2}/{m_{\widetilde{f}_j}^2})\; ,
\end{equation}
with $f=l,u,d$. The neutralino-fermion-sfermion couplings are
\begin{eqnarray}
g_{L\,ij}^{\widetilde{\chi}^0 f \widetilde{f}} \! &=&\!
-\sqrt{2}\, g\, T_3^f\, N_{i2}^* (U^{\widetilde{f}})_{1j}^*
-\sqrt{2}\, g\, t_W\, (Q_f-T_3^f) N_{i1}^* (U^{\widetilde{f}})_{1j}^*
-h_f N_{i\alpha}^* (U^{\widetilde{f}})_{2j}^*\,,\nonumber\\
g_{R\,ij}^{\widetilde{\chi}^0 f \widetilde{f}} \! &=&\!
\sqrt{2}\, g\, t_W\, Q_f\, N_{i1} (U^{\widetilde{f}})_{2j}^*
-h_f^* N_{i\alpha} (U^{\widetilde{f}})_{1j}^*\,,
\end{eqnarray}
where the Higgsino index $\alpha=3\,(f=l,d)$ or $ 4\,(f=u)$, $T_3^{l,d}=-1/2$
and $T_3^{u}=+1/2$ and the loop function is given by
\begin{eqnarray}
B(r) \ =\ \frac{1}{2(1-r)^2}\left(1+r+\frac{2 r \ln{r}}{1-r}\right)\; ,
\end{eqnarray}
with $B(1)=1/6$.
As well as the EDMs, the  neutralino loops can induce
non-vanishing chromo-electric dipole  moments
(CEDMs) for the quarks as follows:
\begin{eqnarray}
\left(d^C_{q=u,d}\right)^{\widetilde\chi^0}&=&
\frac{g_s}{16\pi^2}\sum_{i=1}^5\sum_{j=1}^2 \frac{m_{\widetilde{\chi}^0_i}}{m_{\widetilde{q}_j}^2}\,
\imag[(g_{R\,ij}^{\widetilde{\chi}^0 q \widetilde{q}})^*\,
g_{L\,ij}^{\widetilde{\chi}^0 q \widetilde{q}}]\,
B({m_{\widetilde{\chi}^0_i}^2}/{m_{\widetilde{q}_j}^2})\,.
\end{eqnarray}

\subsection{Two-loop Barr--Zee EDMs}
Beyond  the  one-loop,  we take account of the contributions from
the two-loop Barr--Zee-type  diagrams.
We have considered the the Barr--Zee diagrams
mediated by the $\gamma$-$\gamma$-$H_i^0$ couplings~\cite{Ellis:2008zy}
and the $\gamma$-$H^\pm$-$W^\mp$ and
$\gamma$-$W^\pm$-$W^\mp$ couplings~\cite{Ellis:2010xm}.
The two-loop diagrams mediated by the
$\gamma$-$H^0$-$Z$ couplings~\cite{Giudice:2005rz,Li:2008kz}
have also been included
taking account of the general CP-violating Higgs-boson mixing.
More explicitly, the contribution from the two-loop
Higgs-mediated Barr--Zee-type diagrams can be decomposed into 
four parts:
\begin{equation}
\left(d^E_f\right)^{\rm BZ} =
\left(d^E_f\right)^{\gamma H^0}+ \
\left(d^E_f\right)^{W^\mp H^\pm}+ \
\left(d^E_f\right)^{W^\mp W^\pm}+ \
\left(d^E_f\right)^{Z H^0}
\end{equation}
where
\begin{eqnarray}
(-Q_f)^{-1}\times \left(\frac{d_f^E}{e}\right)^{\gamma H^0} 
\!&=&\! \sum_{q=t,b}\Bigg\{
\frac{3\alpha_{\rm em}\,Q_q^2\,m_f}{32\pi^3}
\sum_{i=1}^5\frac{g^P_{H_i\bar{f}f}}{M_{H_i}^2}
\sum_{j=1,2} g_{H_i\widetilde{q}_j^*\widetilde{q}_j}\,F(\tau_{\widetilde{q}_ji})
\nonumber \\
&&\! +\frac{3\alpha_{\rm em}^2\,Q_q^2\,m_f}{8\pi^2s_W^2M_W^2}
\sum_{i=1}^5\left[
g^P_{H_i\bar{f}f} g^S_{H_i\bar{q}q}\,f(\tau_{qi})
+g^S_{H_i\bar{f}f} g^P_{H_i\bar{q}q}\,g(\tau_{qi})
\right]\Bigg\}  \nonumber \\
&&\! +\
\frac{\alpha_{\rm em}\,m_f}{32\pi^3}
\sum_{i=1}^5\frac{g^P_{H_i\bar{f}f}}{M_{H_i}^2}
\sum_{j=1,2} g_{H_i\widetilde{\tau}_j^*\widetilde{\tau}_j}\,F(\tau_{\widetilde{\tau}_ji})
\nonumber \\
&&\! +\
\frac{\alpha_{\rm em}^2\,m_f}{8\pi^2s_W^2M_W^2}
\sum_{i=1}^5\left[
g^P_{H_i\bar{f}f} g^S_{H_i\tau^+\tau^-}\,f(\tau_{\tau i})
+g^S_{H_i\bar{f}f} g^P_{H_i\tau^+\tau^-}\,g(\tau_{\tau i})
\right] \nonumber \\
&&\! +\ \frac{\alpha_{\rm em}^2\,m_f}{4\sqrt{2}\pi^2s_W^2M_W} \nonumber \\
&&\hspace{0.5cm}\times
\sum_{i=1}^5\sum_{j=1,2}\frac{1}{m_{\chi^\pm_j}}\left[
g^P_{H_i\bar{f}f} g^S_{H_i\chi^+_j\chi^-_j}\,f(\tau_{\chi_j^\pm i})
+g^S_{H_i\bar{f}f} g^P_{H_i\chi^+_j\chi^-_j}\,g(\tau_{\chi_j^\pm i})
\right]\,,
\nonumber \\ \\
\left(\frac{d^E_{f_\downarrow}}{e}\right)^{W^\mp H^\pm} &= &
\frac{\alpha^2}{64\pi^2 s_W^4}\,
\left(\frac{-\sqrt{2}\, g_{f_\uparrow f_\downarrow}}{g}\right)\,
\frac{1}{M_{H^\pm}^2}\,
\nonumber \\
&& \hspace{-2.0cm}
\times \sum_{i=1}^5 \sum_{j=1}^2 \Bigg\{ \int_0^1 {\rm d}x\, \frac{1}{1-x}\,
J\left(r_{WH^\pm},\frac{r_{\widetilde\chi^\pm_jH^\pm}}{1-x}+
\frac{r_{\widetilde\chi^0_iH^\pm}}{x} \right)
\nonumber \\
&& \Bigg[
~\imag{\left((g^S_{H^+\bar{f}_\uparrow f_\downarrow} 
+ i g^P_{H^+\bar{f}_\uparrow f_\downarrow})\,
G^{RL}_+\right)} \,m_{\widetilde\chi_j^\pm}\,x^2
\nonumber \\
&&  +
\imag{\left(( g^S_{H^+\bar{f}_\uparrow f_\downarrow} 
+ i  g^P_{H^+\bar{f}_\uparrow f_\downarrow})\,
G^{LR}_+\right)} \,m_{\widetilde\chi_i^0}\,(1-x)^2
\nonumber \\
&&  +
\imag{\left(( g^S_{H^+\bar{f}_\uparrow f_\downarrow} 
+ i  g^P_{H^+\bar{f}_\uparrow f_\downarrow})\,
G^{RL}_-\right)} \,m_{\widetilde\chi_j^\pm}\,x
\nonumber \\
&&  +
\imag{\left(( g^S_{H^+\bar{f}_\uparrow f_\downarrow} 
+ i  g^P_{H^+\bar{f}_\uparrow f_\downarrow})\,
G^{LR}_-\right)} \,m_{\widetilde\chi_i^0}\,(1-x) \Bigg] \Bigg\}\; ,
\\
\left(\frac{d^E_{f}}{e}\right)^{W^\mp W^\pm} &=&
\frac{\alpha^2}{32\pi^2 s_W^4}\,\sum_{i=1}^5 \sum_{j=1}^2\,
\imag{\left[g^L_{W^+\widetilde\chi_i^0\widetilde\chi_j^-}
\left(g^R_{W^+\widetilde\chi_i^0\widetilde\chi_j^-}\right)^*\right]}\,
\frac{m_f\, m_{\widetilde\chi_i^0}\, m_{\widetilde\chi_j^\pm}}{M_W^4}\,
f_{WW}(r_i,r_j)\; , \nonumber \\
\end{eqnarray}
with $\tau_{xi} =  m_x^2/M_{H_i}^2$,
$r_{xy} \equiv M_x^2/M_y^2$, 
$r_j\equiv m_{\widetilde\chi_j^\pm}^2/M_W^2$ and
$r_i\equiv m_{\widetilde\chi_i^0}^2/M_W^2$ and
$(f_\uparrow,f_\downarrow)=(u,d)\,,(\nu_l\,,l)$.
The $W$-boson couplings to the charginos and neutralinos are given by
\begin{eqnarray}
g^L_{W^+\widetilde\chi_i^0\widetilde\chi_j^-} &=&
N_{i3} (C_L)^*_{j2} +\sqrt{2} N_{i2} (C_L)^*_{j1}\; ,
\nonumber \\
g^R_{W^+\widetilde\chi_i^0\widetilde\chi_j^-} &=&
-N_{i4}^* (C_R)^*_{j2} +\sqrt{2} N_{i2}^* (C_R)^*_{j1}\; \,,
\end{eqnarray}
and, with $A,B=L,R$,
\begin{eqnarray}
\hspace{-0.7cm}
G^{AB}_{\pm} &\equiv & ~
\left(g^S_{H^+\widetilde\chi_i^0\widetilde\chi_j^-}\right)^*
\left(g^A_{W^+\widetilde\chi_i^0\widetilde\chi_j^-} \pm
g^B_{W^+\widetilde\chi_i^0\widetilde\chi_j^-} \right) +
i\,\left(g^P_{H^+\widetilde\chi_i^0\widetilde\chi_j^-}\right)^*
\left(g^A_{W^+\widetilde\chi_i^0\widetilde\chi_j^-} \mp
g^B_{W^+\widetilde\chi_i^0\widetilde\chi_j^-} \right).
\end{eqnarray}
For the loop functions $F(\tau)$, $f(\tau)$, $g(\tau)$, $J(a,b)$, 
and $f_{WW}(r_i\,,r_j)$,
we refer to, for example, Refs.~\cite{Ellis:2008zy,Ellis:2010xm} 
and references therein.
%
Finally, for $(d_f^E)^{Z H^0}$, we take account of the 
dominant fermionic contributions given by
\begin{eqnarray}
\left(\frac{d_f^E}{e}\right)^{Z H^0} &=&
\frac{\alpha^2 v_{Z\bar{f}f}}{16\sqrt{2}\pi^2c_W^2s_W^4}\,
\frac{m_f}{M_W}\,
\sum_{q=t,b} 
\frac{3Q_qm_q}{\sqrt{2}M_W}\,
\nonumber \\ 
&\times & \sum_{i=1}^5 \Bigg[\ 
g^S_{H_i\bar{f}f} \left(v_{Z\bar{q}q} g^P_{H_i\bar{q}q}\right)\,
\frac{m_q}{m_{H_i}^2}\,
\int_0^1 {\rm d}x\frac{1}{x}
J\left(r_{ZH_i},\frac{r_{qH_i}}{x(1-x)}\right)
\nonumber \\ 
&& \hspace{0.6cm}+
g^P_{H_i\bar{f}f} \left(v_{Z\bar{q}q} g^S_{H_i\bar{q}q}\right)\,
\frac{m_q}{m_{H_i}^2}\,
\int_0^1 {\rm d}x\frac{1-x}{x}
J\left(r_{ZH_i},\frac{r_{qH_i}}{x(1-x)}\right) \Bigg]
\nonumber \\
&-& \frac{\alpha^2 v_{Z\bar{f}f}}{16\sqrt{2}\pi^2c_W^2s_W^4}\,
\frac{m_f}{M_W}\,\frac{m_\tau}{\sqrt{2}M_W}\,
\nonumber \\
&\times & \sum_{i=1}^5 \Bigg[\
g^S_{H_i\bar{f}f} \left(v_{Z\tau^+\tau^-} g^P_{H_i\tau^+\tau^-}\right)\,
\frac{m_\tau}{m_{H_i}^2}\,
\int_0^1 {\rm d}x\frac{1}{x}
J\left(r_{ZH_i},\frac{r_{\tau H_i}}{x(1-x)}\right)
\nonumber \\
&& \hspace{0.6cm}+
g^P_{H_i\bar{f}f} \left(v_{Z\tau^+\tau^-} g^S_{H_i\tau^+\tau^-}\right)\,
\frac{m_\tau}{m_{H_i}^2}\,
\int_0^1 {\rm d}x\frac{1-x}{x}
J\left(r_{ZH_i},\frac{r_{\tau H_i}}{x(1-x)}\right) \Bigg]
\nonumber \\
&-& \frac{\alpha^2 v_{Z\bar{f}f}}{16\sqrt{2}\pi^2c_W^2s_W^4}\,
\frac{m_f}{M_W}
\nonumber \\
&\times & \sum_{k=1}^5 \sum_{i,j=1}^2\Bigg[
g^S_{H_k\bar{f}f}\imag\left(
a_{Z\widetilde\chi^+_j\widetilde\chi^-_i} g^S_{H_k\widetilde\chi^+_i\widetilde\chi^-_j}+i\,
v_{Z\widetilde\chi^+_j\widetilde\chi^-_i} g^P_{H_k\widetilde\chi^+_i\widetilde\chi^-_j}\right)
\nonumber \\
&&\hspace{2.3cm}\times\,
\frac{m_{\widetilde\chi^-_j}}{m_{H_k}^2}\,
\int_0^1 {\rm d}x\frac{1}{x}
J\left(r_{ZH_k},
\frac{x r_{\widetilde\chi_i^- H_k}+(1-x) r_{\widetilde\chi_j^- H_k}}{x(1-x)}\right)
\nonumber \\
&&\hspace{1.3cm} +
g^P_{H_k\bar{f}f}\imag\left(
i\,v_{Z\widetilde\chi^+_j\widetilde\chi^-_i} g^S_{H_k\widetilde\chi^+_i\widetilde\chi^-_j}-
a_{Z\widetilde\chi^+_j\widetilde\chi^-_i} g^P_{H_k\widetilde\chi^+_i\widetilde\chi^-_j}\right)
\nonumber \\
&&\hspace{2.3cm}\times\,
\frac{m_{\widetilde\chi^-_j}}{m_{H_k}^2}\,
\int_0^1 {\rm d}x\frac{1-x}{x}
J\left(r_{ZH_k},
\frac{x r_{\widetilde\chi_i^- H_k}+(1-x) r_{\widetilde\chi_j^- H_k}}{x(1-x)}\right)
\Bigg]\,.
\nonumber \\
\end{eqnarray}
The $Z$-boson couplings to the charginos are given by
\begin{eqnarray}
{\cal L}_{Z\widetilde\chi^+\widetilde\chi^-} = -\,g_Z\,
\overline{\widetilde\chi^-_i}\,\gamma^\mu\,
\left(v_{Z\chi^+_i\widetilde\chi^-_j} - a_{Z\chi^+_i\widetilde\chi^-_j} \gamma_5\right)\, 
\widetilde\chi^-_j\,Z_\mu
\end{eqnarray}
where $g_Z=g/c_W=e/(s_Wc_W)$ and
\begin{eqnarray}
v_{Z\chi^+_i\widetilde\chi^-_j}& = &
\frac{1}{4}\left[
\left(C_L\right)_{i2} \left(C_L\right)_{j2}^* +
\left(C_R\right)_{i2} \left(C_R\right)_{j2}^*
\right] \ - \ c_W^2 \,\delta_{ij}\,,
\nonumber \\
a_{Z\chi^+_i\widetilde\chi^-_j} & = &
\frac{1}{4}\left[
\left(C_L\right)_{i2} \left(C_L\right)_{j2}^* -
\left(C_R\right)_{i2} \left(C_R\right)_{j2}^*
\right]\,.
\end{eqnarray}
And the $Z$-boson couplings to the quarks and leptons are given by
\begin{eqnarray}
{\cal L}_{Z\bar{f}f} = -\,g_Z\,
\bar{f}\,\gamma^\mu\,\left(v_{Z\bar{f}f} - a_{Z\bar{f}f} \gamma_5\right)\,f\,Z_\mu
\end{eqnarray}
with $v_{Z\bar{f}f}=T_{3L}^f/2-Q_f s_W^2$ and $a_{Z\bar{f}f}=T_{3L}^f/2$.
For the SM quarks and leptons,
$T_{3L}^{\,u, \nu}=+1/2$ and $T_{3L}^{\,d, e}=-1/2$.

In addition to EDMs, the two-loop  Higgs-mediated
Barr-Zee graphs also generate CEDMs
of the light quarks $q_l=u,d$ which take the forms:
\begin{eqnarray}
\left(d_{q_l}^C\right)^{\rm BZ} \!&=&\! - \sum_{q=t,b}\Bigg\{
\frac{g_s\,\alpha_s\,m_{q_l}}{64\pi^3}
\sum_{i=1}^5\frac{g^P_{H_i\bar{q}_lq_l}}{M_{H_i}^2}
\sum_{j=1,2} g_{H_i\widetilde{q}_j^*\widetilde{q}_j}\,F(\tau_{\widetilde{q}_ji})
\nonumber \\
&&~~ +\frac{g_s\,\alpha_s\,\alpha_{\rm em}\,m_{q_l}}{16\pi^2s_W^2M_W^2}
\sum_{i=1}^5\left[
g^P_{H_i\bar{q}_lq_l} g^S_{H_i\bar{q}q}\,f(\tau_{qi})
+g^S_{H_i\bar{q}_lq_l} g^P_{H_i\bar{q}q}\,g(\tau_{qi})
\right]\Bigg\}\,.  
\end{eqnarray}

\subsection{Observable EDMs}
In this subsection, we briefly review the dependence of the
Thallium, neutron, Mercury, deuteron, and Radium EDMs on the 
(C)EDMs of quarks and leptons
and the coefficients of  the dimension-six
Weinberg operator and the four-fermion operators.

\subsubsection{Thallium EDM}
The Thallium EDM receives   contributions   mainly   from   two
terms~\cite{KL,Pospelov:2005pr}:
\begin{equation}
d_{\rm Tl}\,[e\,{\rm cm}]\ =\ -585\cdot d_e^E\,[e\,{\rm cm}]\:
-\: 8.5\times 10^{-19}\,[e\,{\rm cm}]\cdot (C_S\,{\rm TeV}^2)\: +\ \cdots\,,
\end{equation}
where $d^E_e$ is the electron EDM and $C_S$ is the coefficient of
the CP-odd electron-nucleon interaction 
${\cal L}_{C_S}=C_S\,\bar{e}i\gamma_5\,e \bar{N}N$ which is given by
\begin{equation}
C_S = C_{de}\frac{29\,{\rm MeV}}{m_d}
+ C_{se}\frac{\kappa\times 220\,{\rm MeV}}{m_s}
+ (0.1\,{\rm GeV})\,
\frac{m_e}{v^2}
\sum_{i=1}^3 \frac{g^S_{H_igg}g^P_{H_i\bar{e}e}}{M_{H_i}^2}
\end{equation}
with $\kappa\equiv\langle N |
m_s \bar{s} s | N \rangle/220~{\rm MeV} \simeq 0.50\pm0.25$
and 
\begin{equation}
g^S_{H_igg}=\sum_{q=t,b}\left\{\frac{2\,x_q}{3}g_{H_i\bar{q}q}^S-\frac{v^2}{12}
\sum_{j=1,2} \frac{g_{H_i\widetilde{q}_j^*\widetilde{q}_j}}{m_{\widetilde{q}_j}^2}
\right\}\,,
\end{equation}
with $x_t=1$ and $x_b=1-0.25 \kappa$.

\subsubsection{Neutron EDM}
For the neutron EDM, we  consider three different hadronic
approaches:  (i)~the  Chiral Quark  Model~(CQM),
(ii)~the Parton Quark Model~(PQM) and (iii)~the QCD sum-rule
technique.
\begin{itemize}
\item In the CQM approach,  the neutron EDM is given by
\begin{eqnarray}
  \label{dnCQM}
d_n \!&=&\! \frac{4}{3}\,d^{\rm NDA}_d\: -\:
                           \frac{1}{3}\,d^{\rm NDA}_u\,, \nonumber \\
d^{\rm NDA}_{q=u,d} \!&=&\! \eta^E\,d_q^E\: +\: \eta^C\,\frac{e}{4\pi}\,d_q^C
\: +\: \eta^G\,\frac{e\Lambda}{4\pi}\,d^{\,G}\,, 
\end{eqnarray}
where  the chiral  symmetry breaking  scale $\Lambda  \simeq 1.19~{\rm
GeV}$ and the $\eta^{E,C,G}$ account for
the renormalization-group (RG) evolution of $d^{E,C}_q$ and $d^G$ from
the electroweak  (EW) scale to the  hadronic   scale.   
For the   QCD  correction  factors we are taking
$\eta^E \simeq 1.53$ and $\eta^C \simeq \eta^G \simeq 3.4$~\cite{Ibrahim:1997gj}.
We note that the  EDM operators $d^{E,C}_q$ and $d^G$ in~(\ref{dnCQM})
are computed at the EW scale.
\item In the PQM approach,  the neutron EDM is given by~\cite{Ellis:1996dg}
\begin{eqnarray}
  \label{dnPQM}
d_n  \!&=&\! \eta^E\, (\Delta^{\rm PQM}_d\,d_d^E\: +\:
\Delta^{\rm PQM}_u\,d_u^E+\Delta^{\rm PQM}_s\,d_s^E)\;,
\end{eqnarray}
with 
$\Delta^{\rm PQM}_d =0.746$,
$\Delta^{\rm PQM}_u =-0.508$, and
$\Delta^{\rm PQM}_s =-0.226$.
\item Using the QCD sum-rule technique,
the neutron EDM is given by~\cite{qcdsumrule1,qcdsumrule2,Demir:2002gg,Demir:2003js,
Olive:2005ru}
\begin{eqnarray}
  \label{dnQCD}
d_n \!&=&\! d_n(d_q^E\,,d_q^C)\: +\: d_n(d^{\,G})\: +\: d_n(C_{bd})\:
+\: \cdots\,, \nonumber \\
d_n(d_q^E\,,d_q^C) \!&=&\! (1.4\pm 0.6)\,(d_d^E-0.25\,d_u^E)\:
+\: (1.1\pm 0.5)\,e\,(d_d^C+0.5\,d_u^C)/g_s\,,
\nonumber \\
d_n(d^{\,G}) \!&\sim&\! \pm\, e\, (20\pm 10)~{\rm MeV} \,d^{\,G}\,,
\nonumber \\
d_n (C_{bd}) \!&\sim &\! \pm\, e\, 2.6\times 10^{-3}~{\rm GeV}^2\,
\left[\frac{C_{bd}}{m_b}\: +\: 0.75\frac{C_{db}}{m_b}\right]\; ,
\end{eqnarray}
where  $d_q^E$ and $d_q^C$  should be  evaluated at  the EW  scale and
$d^{\,G}$  at  the 1  GeV  scale,
$d^G\big|_{1~{\rm  GeV}} \simeq (\eta^G/0.4)\, d^G\big|_{\rm
EW}  \simeq 8.5\,  d^G\big|_{\rm EW}$~\cite{Demir:2002gg}.
In the numerical estimates 
we take the positive signs for $d_n(d^{\,G})$ and $d_n (C_{bd})$.
\end{itemize}

\subsubsection{Mercury EDM}
Using the QCD sum rules~\cite{Demir:2003js,Olive:2005ru}, we estimate
the Mercury EDM as follows:
\begin{eqnarray}
d^{\rm \,I\,,II\,,III\,,IV}_{\rm Hg} \!&=&\!
d^{\rm \,I\,,II\,,III\,,IV}_{\rm Hg}[S]
+10^{-2} d_e^E 
+(3.5\times 10^{-3} {\rm GeV})\,e\,C_S 
\nonumber \\ \!&&\!
+\ (4\times 10^{-4}~{\rm GeV})\,e\,\left[C_P+
\left(\frac{Z-N}{A}\right)_{\rm Hg}\,C^\prime_P\right]\,,
\end{eqnarray}
where $d^{\rm \,I\,,II\,,III\,,IV}_{\rm Hg}[S]$ 
denotes the Mercury EDM induced by the Schiff moment.
The parameters $C_P$ and $C^\prime_P$  are the couplings of 
electron-nucleon interactions as in
${\cal L}_{C_P}\ =\ C_P\,\bar{e}e\,\bar{N}i\gamma_5 N\: +\:
C^\prime_P\,\bar{e}e\,\bar{N}i\gamma_5 \tau_3  N$ and they are given by
\cite{Ellis:2008zy}
\begin{eqnarray}
C_P & \simeq &
-\,375~{\rm MeV}\,\sum_{q=c,s,t,b} \frac{C_{eq}}{m_q}\,,\nonumber \\
C^\prime_P & \simeq\ & -\,806~{\rm MeV}\,\frac{C_{ed}}{m_d}\,
-\,181~{\rm MeV}\,\sum_{q=c,s,t,b} \frac{C_{eq}}{m_q}\,.
\end{eqnarray}
We take account  of  the uncertainties in the calculation of
the Schiff-moment induced Mercury EDM
as follows~\cite{Ellis:2011hp}:
\begin{eqnarray}
d^{\rm \,I}_{\rm Hg}[S]\ & \simeq\ & 
1.8 \times 10^{-3}\, e\,\bar{g}^{(1)}_{\pi NN}\,/{\rm GeV}\,,
\nonumber \\
d^{\rm \,II}_{\rm Hg}[S]\ & \simeq\ &
7.6 \times 10^{-6}\, e\,\bar{g}^{(0)}_{\pi NN}\,/{\rm GeV}+
1.0 \times 10^{-3}\, e\,\bar{g}^{(1)}_{\pi NN}\,/{\rm GeV}\,,
\nonumber \\
d^{\rm \,III}_{\rm Hg}[S]\ & \simeq\ &
1.3 \times 10^{-4}\, e\,\bar{g}^{(0)}_{\pi NN}\,/{\rm GeV}+
1.4 \times 10^{-3}\, e\,\bar{g}^{(1)}_{\pi NN}\,/{\rm GeV}\,,
\nonumber \\
d^{\rm \,IV}_{\rm Hg}[S]\ & \simeq\ &
3.1 \times 10^{-4}\, e\,\bar{g}^{(0)}_{\pi NN}\,/{\rm GeV}+
9.5 \times 10^{-5}\, e\,\bar{g}^{(1)}_{\pi NN}\,/{\rm GeV}\,.
\end{eqnarray}
where
\begin{eqnarray}
\bar{g}^{(0)}_{\pi NN} \!&=&\! 
0.4\times 10^{-12}\,\frac{(d_u^C+d_d^C)/g_s}{10^{-26}{\rm cm}}\,
\frac{|\langle \bar{q} q\rangle |}{(225\,{\rm MeV})^3}\,,
\nonumber \\
\bar{g}^{(1)}_{\pi NN} \!&=&\!
2^{+4}_{-1}\times 10^{-12}\,\frac{(d_u^C-d_d^C)/g_s}{10^{-26}{\rm cm}}\,
\frac{|\langle \bar{q} q\rangle |}{(225\,{\rm MeV})^3} \nonumber \\
&& -\, 8\times 10^{-3} {\rm GeV}^3\,
\left[\frac{0.5C_{dd}}{m_d}+3.3\kappa\frac{C_{sd}}{m_s}
+(1-0.25\kappa)\frac{C_{bd}}{m_b} \right]\,.
\end{eqnarray}

\subsubsection{Deuteron EDM}
For the deuteron EDM, we use~\cite{Lebedev:2004va,Ellis:2008zy}:
\begin{eqnarray}
d_{D} \!&\simeq &\! -\left[5^{+11}_{-3} + (0.6\pm 0.3)
  \right]\,e\,(d^C_u-d^C_d)/g_s
\nonumber \\
\!&&\! -(0.2\pm 0.1)\,e\,(d^C_u+d^C_d)/g_s\:
+\: (0.5 \pm 0.3) (d^E_u+d^E_d) \nonumber \\
\!&&\! +(1\pm 0.2)\times 10^{-2}\,e\,{\rm GeV}^2\,
\left[\frac{0.5C_{dd}}{m_d}+3.3\kappa\frac{C_{sd}}{m_s}
+(1-0.25\kappa)\frac{C_{bd}}{m_b} \right] \nonumber \\
&& \pm\ e\, (20\pm 10)~{\rm MeV} \,d^{\,G}\; .
\end{eqnarray}
In  the above,  $d^{\,G}$  is evaluated  at the  $1$ GeV scale,
and the coupling coefficients $g_{d,s,b}$ appearing in $C_{dd,sd,bd}$
are computed at energies 1~GeV, 1~GeV and $m_b$, respectively.
All other  EDM  operators  are calculated  at  the EW  scale.
In the numerical estimates 
we take the positive sign for $d^{\,G}$.

\subsubsection{Radium EDM}
For the EDM of $^{225}$Ra,
we use~\cite{Ellis:2011hp}:
\begin{eqnarray}
d_{\rm Ra} \simeq 
d_{\rm Ra}[S] \simeq 
-8.7 \times 10^{-2}\, e\,\bar{g}^{(0)}_{\pi NN}\,/{\rm GeV}
+3.5 \times 10^{-1}\, e\,\bar{g}^{(1)}_{\pi NN}\,/{\rm GeV}\,.
\end{eqnarray}
We  note that  the $\bar{g}^{(1)}_{\pi  NN}$ contribution  to
the  Radium EDM is about  200 times larger than that to the 
Mercury EDM $d^{\rm \,I}_{\rm Hg}[S]$~\cite{Engel:2003rz}.

\section{Numerical Analysis}

\begin{figure}[t!]
\begin{center}
\includegraphics[width=14.3cm]{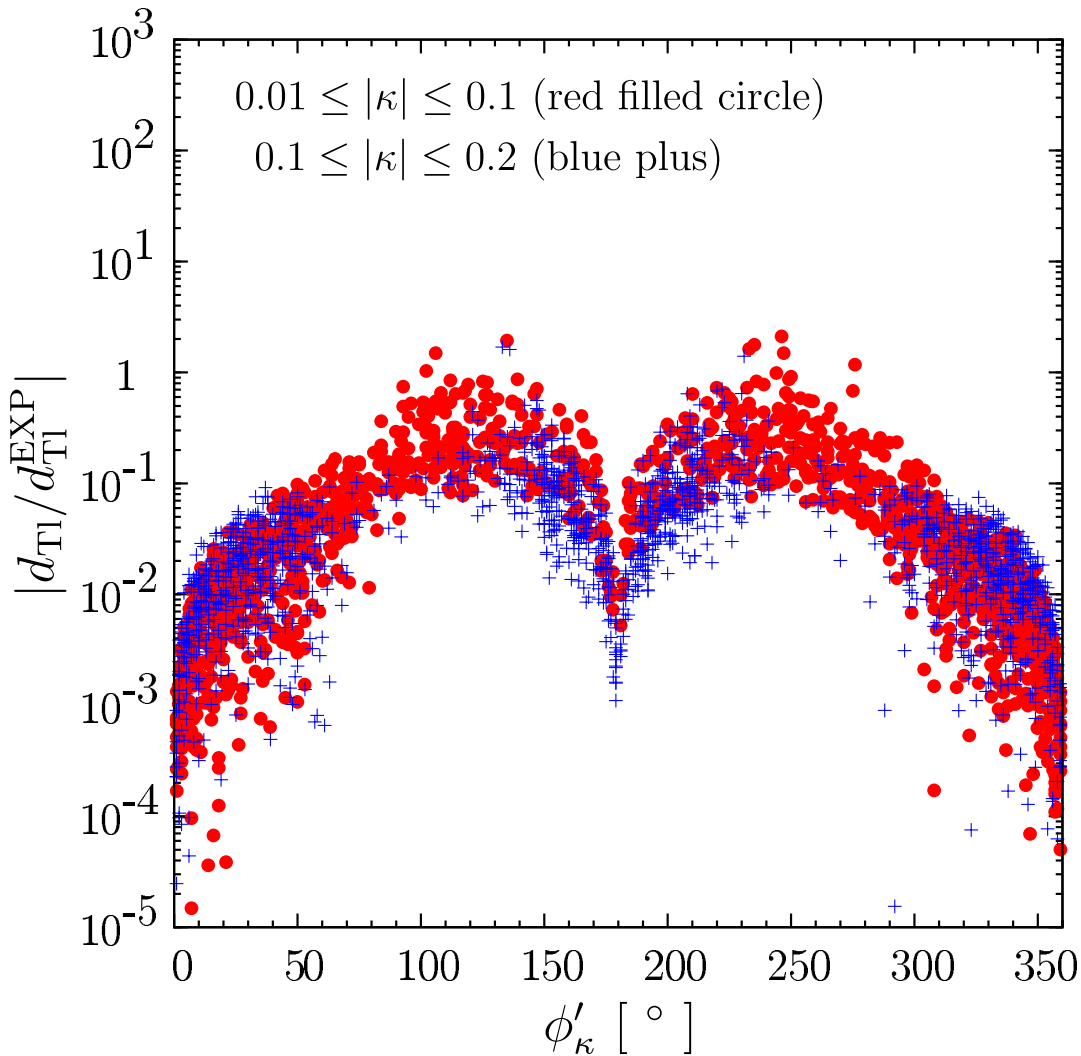}
\end{center}
\caption{The absolute value of the Thallium EDM $d_{\rm Tl}$
divided by the current experimental limit $d_{\rm Tl}^{\rm EXP}$
as a function of $\phi_\kappa^\prime$ varying
$|\lambda|$, $|\kappa|$, $|A_\lambda|$, and $|A_\kappa|$
over the ranges given by Eq.~(\ref{eq:ranges})
for the scenario specified by Eq.~(\ref{eq:scenario}).
Especially, the $|\kappa|$ range is divided into 2 regions:
$0.01\leq |\kappa| < 0.1$ (red filled circle), and
$0.1\leq |\kappa| \leq 0.2$ (blue plus) .
}
\label{fig:Tl_scatter}
\end{figure}
\begin{figure}[t!]
\begin{center}
\includegraphics[width=5.3cm]{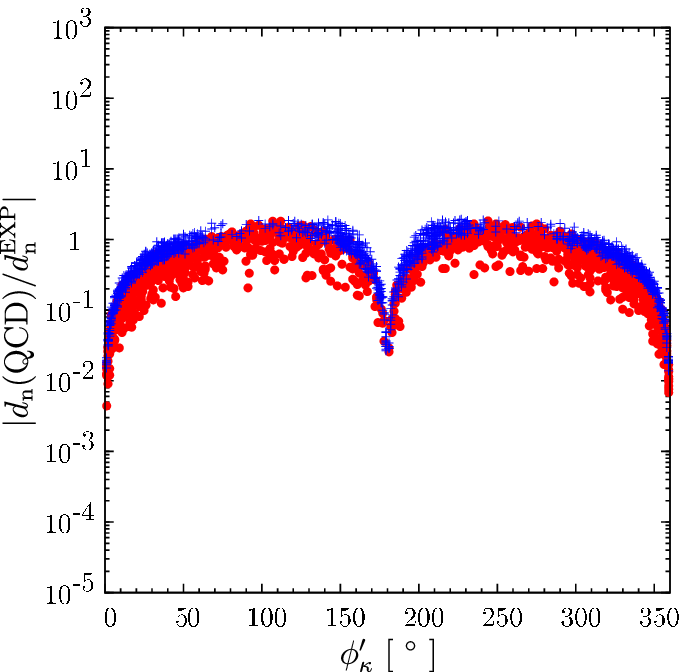}
\includegraphics[width=5.3cm]{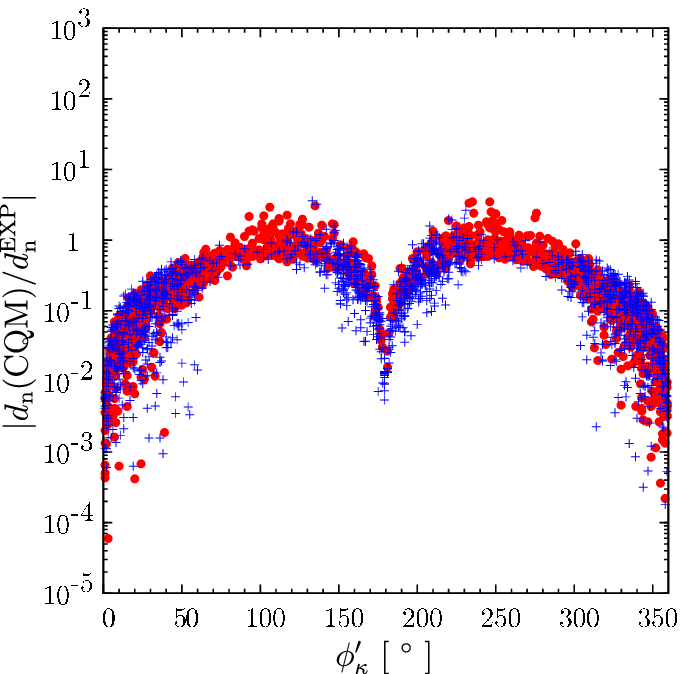}
\includegraphics[width=5.3cm]{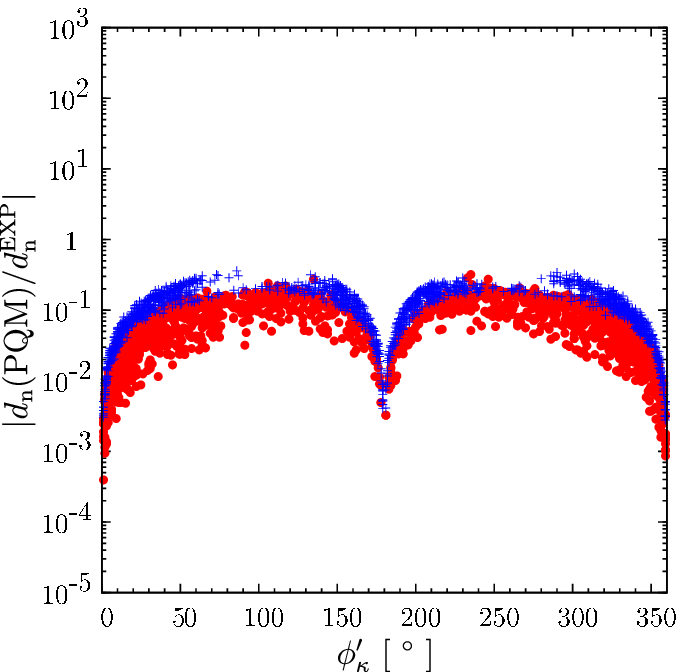}
\end{center}
\caption{The same as in Fig.~\ref{fig:Tl_scatter} but
for the neutron EDM in the QCD sum rules (left),
CQM (middle) and PQM approaches (right).
}
\label{fig:n_scatter}
\end{figure}
\begin{figure}[t!]
\begin{center}
\includegraphics[width=7.3cm]{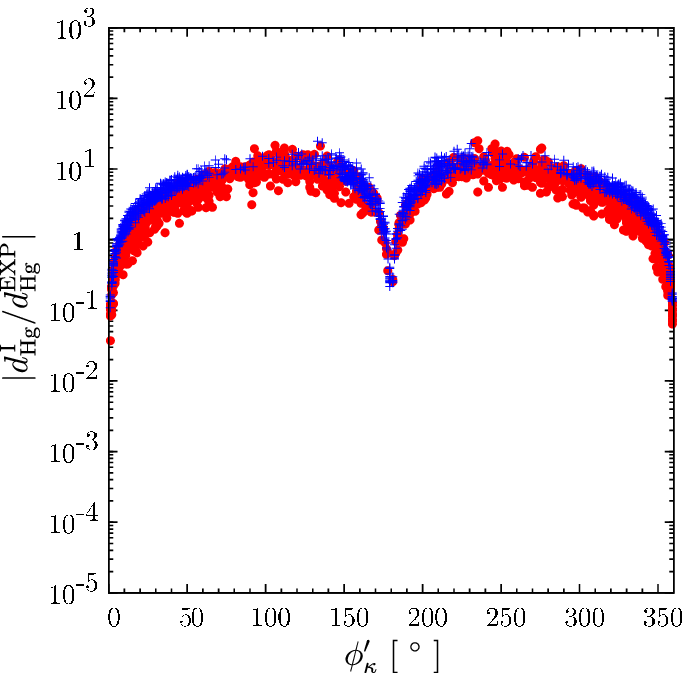}
\includegraphics[width=7.3cm]{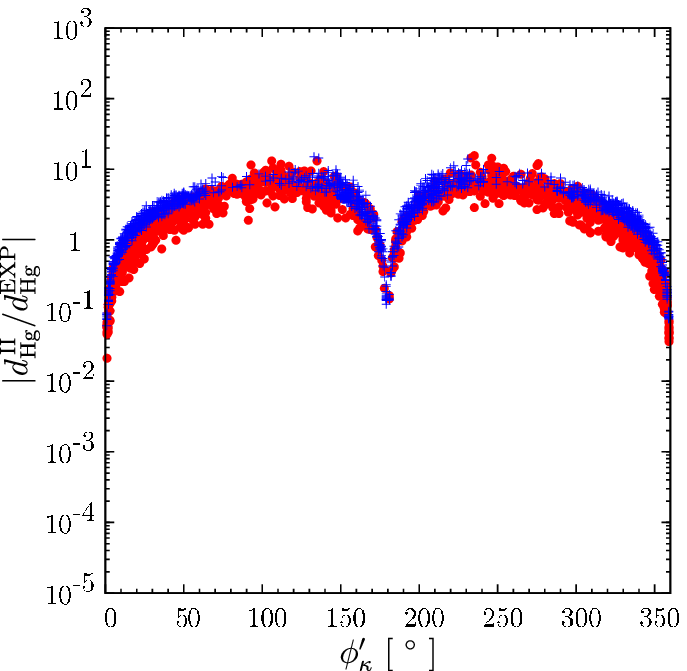}
\includegraphics[width=7.3cm]{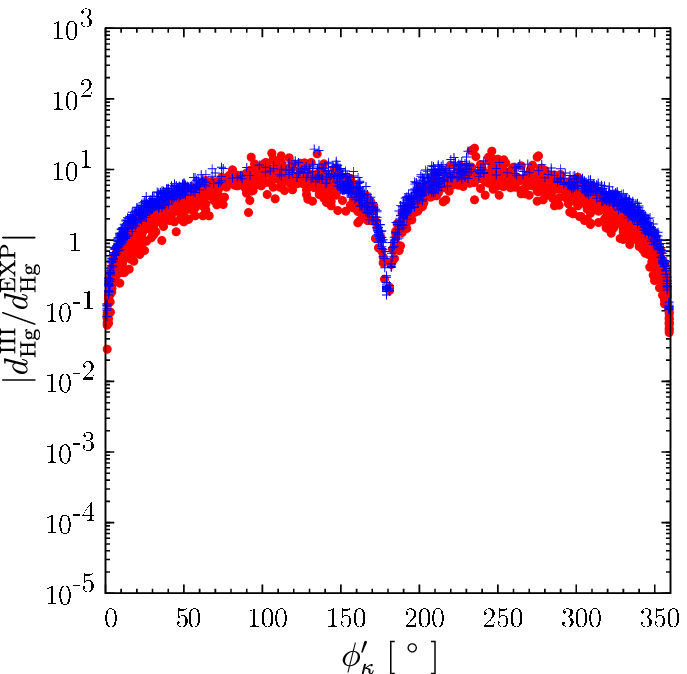}
\includegraphics[width=7.3cm]{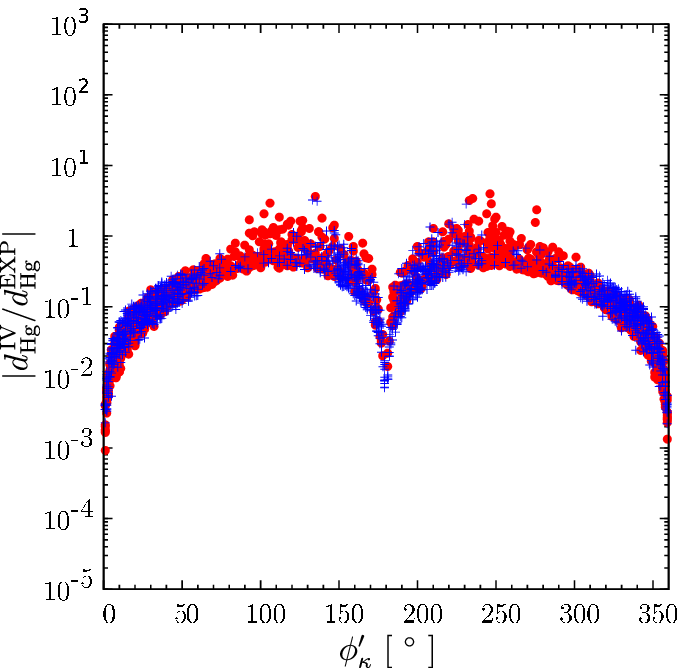}
\end{center}
\caption{The same as in Fig.~\ref{fig:Tl_scatter} but
for the Mercury EDM:
$d^{\rm\,I}_{\rm Hg}$ (upper left),
$d^{\rm\,II}_{\rm Hg}$ (upper right),
$d^{\rm\,III}_{\rm Hg}$ (lower left), and
$d^{\rm\,IV}_{\rm Hg}$ (lower right).
}
\label{fig:Hg_scatter}
\end{figure}

The scenario we are considering has an intermediate value of $\tan\beta$ with
small $v_S \sim v$:
\begin{eqnarray}
&&
\tan\beta=5\,, \ \
v_S=200~{\rm GeV}\,,
\nonumber \\
&&
M_{{\widetilde Q}_{1,2,3}}=M_{{\widetilde U}_{1,2,3}}=M_{{\widetilde D}_{1,2,3}}=
M_{{\widetilde L}_{1,2,3}}=M_{{\widetilde E}_{1,2,3}}=1~{\rm TeV}\,,
\nonumber \\
&&
|A_e|=|A_u|=|A_d|=|A_s|=
|A_\tau|=|A_t|=|A_b|=1~{\rm TeV}\,,
\nonumber \\
&&
\phi_{A_e}=\phi_{A_u}=\phi_{A_d}=\phi_{A_s}=
\phi_{A_\tau}=\phi_{A_t}=\phi_{A_b}=0\,;\ 
\phi_{1,2}=\pi\,,
\phi_{3}=0\,,
\nonumber \\
&&
\phi_\lambda^\prime=0\,;  \ \
{\rm sign}\,[\cos(\phi^\prime_\kappa+\phi_{A_\kappa})] =
{\rm sign}\,[\cos(\phi^\prime_\lambda+\phi_{A_\lambda})]  = + 1 \,,
\label{eq:scenario}
\end{eqnarray}
while varying
\begin{equation}
|\lambda|\,,|\kappa|\,; \ \
|A_\lambda|\,,|A_\kappa| \,; \ \
\phi_\kappa^\prime\,.
\end{equation}
We have chosen $M_1=M_2=-200$ GeV to fix the neutralino sector
\footnote{In some regions of the parameter space, we find that
$\Gamma(Z\to\widetilde\chi^0_1\widetilde\chi^0_1)$ becomes sizable around
$\phi_\kappa^\prime = 180^\circ$, violating the LEP bound
on the non-SM contributions to the invisible $Z$ decay width, 
$\Delta\Gamma_{\rm inv}<2$ MeV~\cite{Hebbeker:1999pi}.
In this section, 
we are presenting our results without including the bound
on $\Delta\Gamma_{\rm inv}$
since it can be easily satisfied  for other choices of 
$M_{1,2}$ without affecting the numerical results much.
}.
We find that a first-order phase transition could occur in some
regions of the parameter space of this scenario~\cite{Funakubo:2005pu}
which is needed for the EWBG~\cite{ewbg}.

In Figs.~\ref{fig:Tl_scatter},~\ref{fig:n_scatter},
and ~\ref{fig:Hg_scatter},
we show the absolute values of
\begin{eqnarray}
&&
d_{\rm Tl}/d_{\rm Tl}^{\rm EXP}\,, \ \
d_{\rm n}/d_{\rm n}^{\rm EXP}\,, \ \
d_{\rm Hg}/d_{\rm Hg}^{\rm EXP}\,,  
\end{eqnarray}
with the current experimental bounds
\begin{eqnarray}
&&
d_{\rm Tl}^{\rm EXP}=9\times 10^{-25}\, e\,{\rm cm}\,, \ \
d_{\rm n}^{\rm EXP}=2.9\times 10^{-26}\, e\,{\rm cm}\,, \ \
d_{\rm Hg}^{\rm EXP}=3.1\times 10^{-29}\, e\,{\rm cm}\,.
\end{eqnarray}
For a given value of $\phi^\prime_\kappa$,
we perform a scan by sampling the four model parameters
in the following ranges:
\begin{eqnarray}
0.75 \leq & |\lambda| & \leq 0.95 \,, \nonumber \\
0.01 \leq & |\kappa| & \leq 0.2 \,, \nonumber \\
|A_\lambda|_{\rm MIN} \leq & |A_\lambda| & \leq 800~{\rm GeV} \,, \nonumber \\
|A_\kappa|_{\rm MIN} \leq & |A_\kappa| & \leq 200~{\rm GeV} \,,
\label{eq:ranges}
\end{eqnarray}
where $|A_{\lambda\,,\kappa}|_{\rm MIN}$ are determined by the 
tadpole conditions~\cite{Cheung:2010ba}.
The lower limit on $|\lambda|$ comes from the chargino mass limit
and that on $|\kappa|$ is derived 
from the global minimum condition and the requirement
of the strong enough first-order electroweak phase transition.
The upper limits on $|\lambda|$ and $|\kappa|$ come from requiring that
there is no serious breakdown of perturbativity below the GUT scale.
Especially, we have taken $0.95$ as the upper bound for $|\lambda|$ 
as done in Ref.~\cite{Funakubo:2005pu}, though it is somewhat 
larger than the usual perturbativity bound ($\sim 0.8$) quoted 
in the NMSSM.  Such a value of $|\lambda|$ might be comfortably
accommodated in, for example, the SUSY fat Higgs 
model which includes a new gauge interaction that becomes
strong at an intermediate scale~\cite{Harnik:2003rs}.
And then the LEP limits, the global minimum condition, and the positivity of
the square of the Higgs-boson mass have been imposed
as described in Ref.~\cite{Cheung:2010ba}.
We note that the region of $|A_\lambda|$ is 
around $|\lambda| v_S \tan\beta/\sqrt{2} \sim 600$ GeV, which
determines the typical masses of the heavier Higgs bosons.

The Thallium EDM is below the current experimental limits over the
whole range of the parameters except for a few 
points around $\phi_\kappa^\prime = 110^\circ$ and $250^\circ$,
see Fig.~\ref{fig:Tl_scatter}. Also, we see that
the ratio $|d_{\rm Tl}/d_{\rm Tl}^{\rm EXP}|$
does not exceed 3. 

Fig.~\ref{fig:n_scatter} shows the neutron EDM in the three 
different approaches.
The estimations in the QCD sum rules and CQM approaches give more or less 
similar results which do not exceed 3 times the current experimental 
limit while the PQM prediction always lies below it.
We observe that larger values of $|\kappa|$ 
lead to larger EDMs in the QCD and PQM cases.

Fig.~\ref{fig:Hg_scatter} shows the four different calculations of
the Mercury EDM. Using the three calculations $d^{\rm\,I}_{\rm Hg}$,
$d^{\rm\,II}_{\rm Hg}$, and $d^{\rm\,III}_{\rm Hg}$, we observe that 
$|d_{\rm Hg}/d_{\rm Hg}^{\rm EXP}|$ can be as large as $\sim 20$ and
so only the small angle region with
$\Delta \phi_\kappa^\prime \sim \pm 10^\circ\,(30^\circ)$ around
$0^\circ$ is allowed when $|\kappa|\geq (<) 0.1$. The small angle region
around $180^\circ$ is also allowed.
While if we adopt $d^{\rm\,IV}_{\rm Hg}$, the Mercury EDM 
does not exceed 4 times the current experimental limit and
much larger values of $\phi_\kappa^\prime$ are still allowed.

To summarize, the tree-level CP phase $\phi_\kappa^\prime$
is hardly constrained by the non-observation  of  
Thallium and neutron EDMs. The Mercury EDM constraint could
be stronger but there is still a room to have large
$\phi_\kappa^\prime \sim 90^\circ$ 
after taking account  of  the uncertainties in the calculation of
the Schiff-moment induced Mercury EDM, $d_{\rm Hg}[S]$.

\begin{figure}[t!]
\begin{center}
\includegraphics[width=7.3cm]{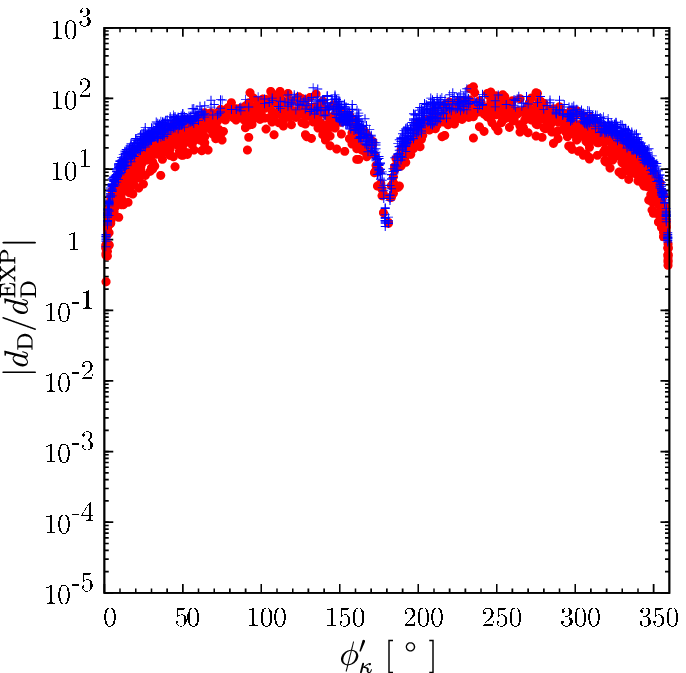}
\includegraphics[width=7.3cm]{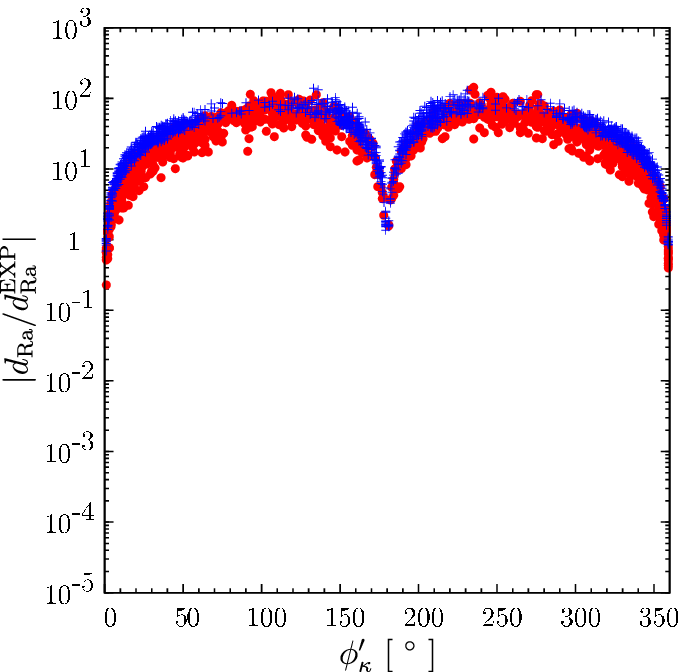}
\end{center}
\caption{The same as in Fig.~\ref{fig:Tl_scatter} but
for the deuteron (left) and Radium (right) EDMs.
}
\label{fig:D_Ra_scatter}
\end{figure}

At this stage, it would be interesting to see whether the proposed
future EDM experiments can probe the CP phase $\phi_\kappa^\prime$.
In this work,
we consider the deuteron EDM and the $^{225}$Ra EDM. The latter
is known to be enhanced  by the  Schiff moment  induced  
by the  presence of  nearby parity-doublet states~\cite{Engel:2003rz}.
In Fig.~\ref{fig:D_Ra_scatter},
we show the absolute values of
\begin{eqnarray}
d_{\rm D}/d_{\rm D}^{\rm EXP}\,   \ \ \ {\rm and} \ \ \
d_{\rm Ra}/d_{\rm Ra}^{\rm EXP}\,.
\end{eqnarray}
For the normalization of the deuteron EDM, we have taken
the projected experimental sensitivity~\cite{Semertzidis:2003iq}:
$d_{\rm D}^{\rm EXP}=3\times 10^{-27}\, e\,{\rm cm}$.
For the Radium EDM, we have taken 
$d_{\rm Ra}^{\rm EXP}=1\times 10^{-27}\, e\,{\rm cm}$.
The chosen value for $d_{\rm Ra}^{\rm EXP}$
is near to a sensitivity which can be achieved in 
one day of  data-taking~\cite{Willmann}.
From Fig.~\ref{fig:D_Ra_scatter}, we see that 
$|d_{\rm D}/d_{\rm D}^{\rm EXP}|$ and 
$|d_{\rm Ra}/d_{\rm Ra}^{\rm EXP}|$
can be as large as $\sim 150$ and 
the larger values of $|\kappa|$ lead to the larger EDMs. 
We observe that both of the experiments have
powerful potential to probe almost the whole region of 
$\phi_\kappa^\prime$.

\begin{figure}[t!]
\begin{center}
\includegraphics[width=14.3cm]{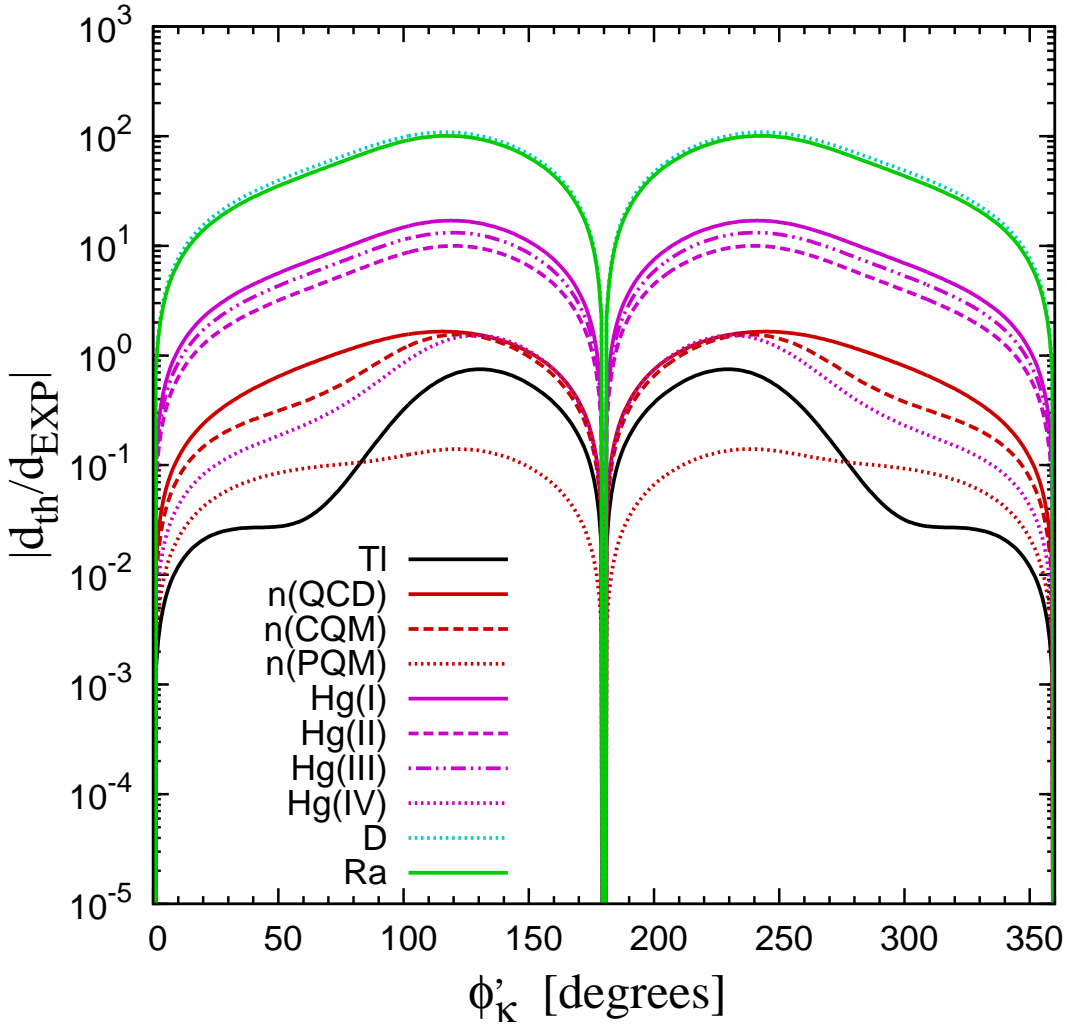}
\end{center}
\caption{The observable EDMs taking
$|\lambda|=0.81$,
$|\kappa|=0.08$,
$|A_\lambda|=575$ GeV, and
$|A_\kappa|=110$ GeV.
The other parameters are fixed as in Eq.~(\ref{eq:scenario})
}
\label{fig:S4}
\end{figure}
\begin{figure}[t!]
\begin{center}
\includegraphics[width=7.3cm]{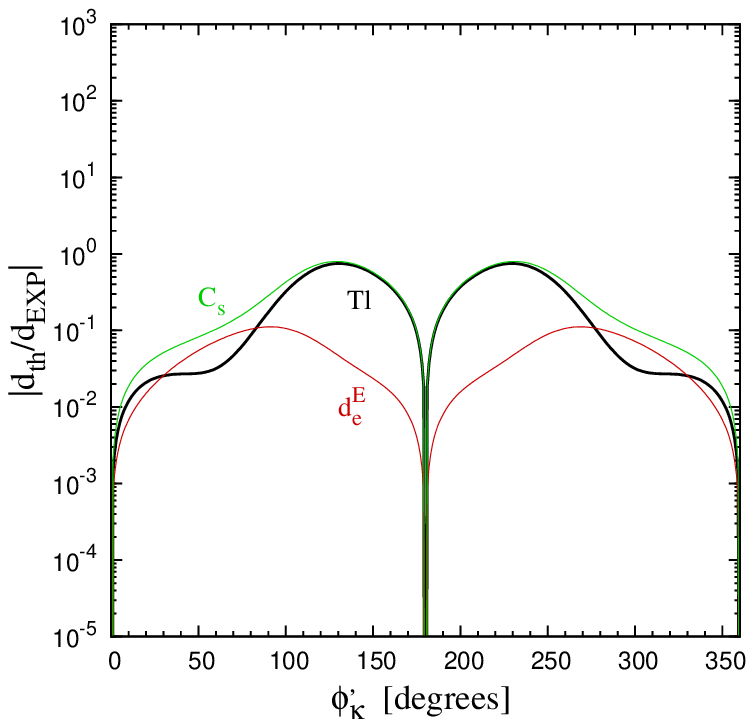}
\includegraphics[width=7.3cm]{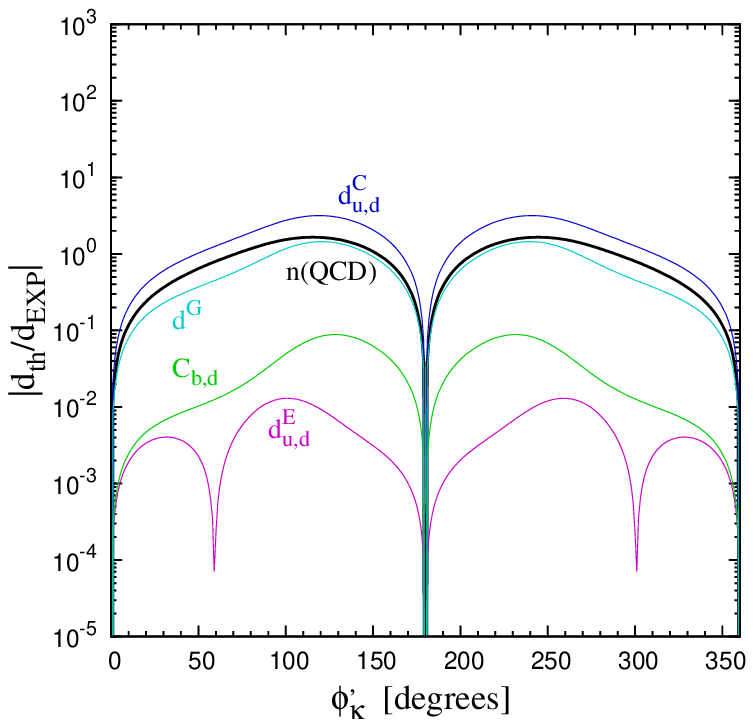}
\includegraphics[width=7.3cm]{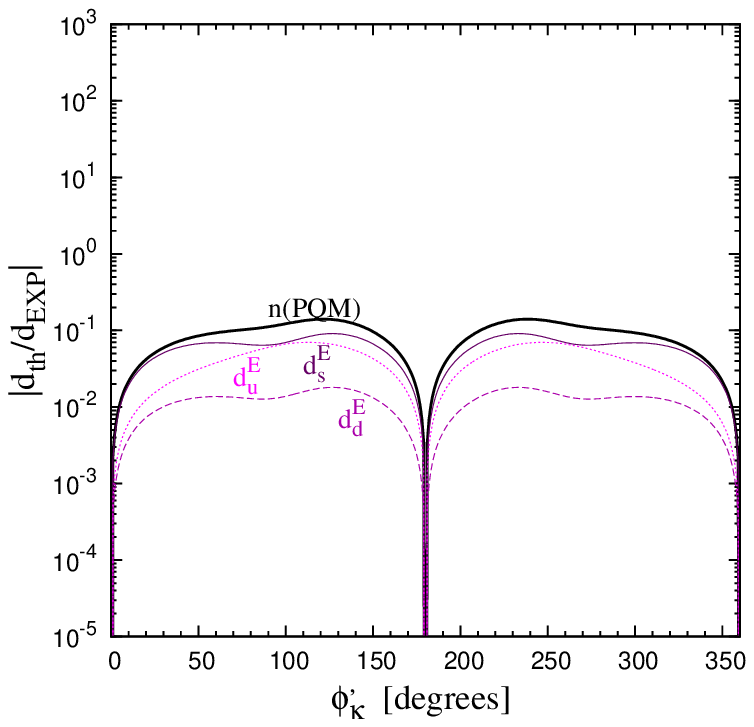}
\includegraphics[width=7.3cm]{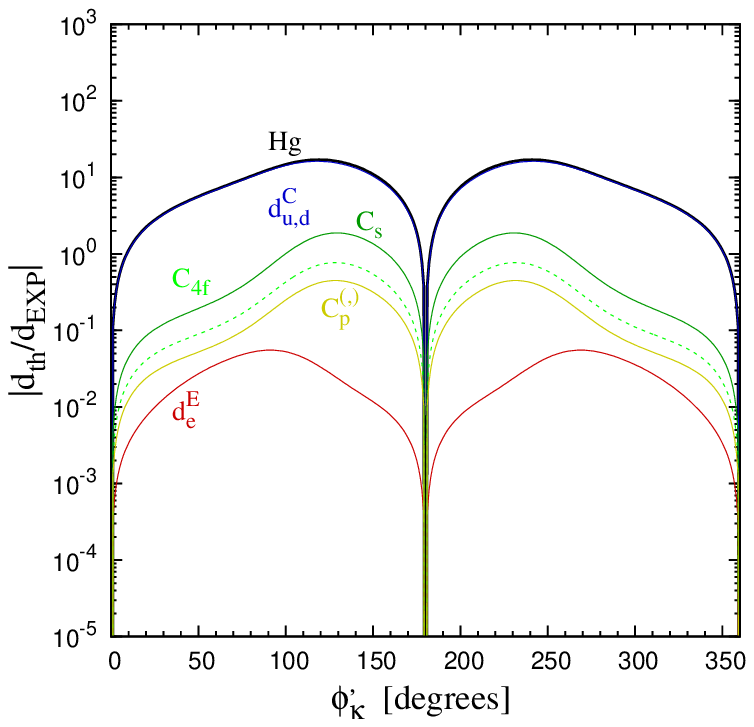}
\end{center}
\caption{The observable EDMs together with the 
constituent contributions taking $|\lambda|=0.81$,
$|\kappa|=0.08$,
$|A_\lambda|=575$ GeV, and
$|A_\kappa|=110$ GeV.
The other parameters are fixed as in Eq.~(\ref{eq:scenario}).
Each frame shows:
the Thallium EDM (upper left),
the neutron EDM in the QCD sum-rule approach (upper right),
the neutron EDM in the PQM (lower left), and 
the Mercury EDM $d^{\rm\,I}_{\rm Hg}$ (lower right).}
\label{fig:S4_Tl_n_Hg}
\end{figure}
\begin{figure}[t!]
\begin{center}
\includegraphics[width=7.3cm]{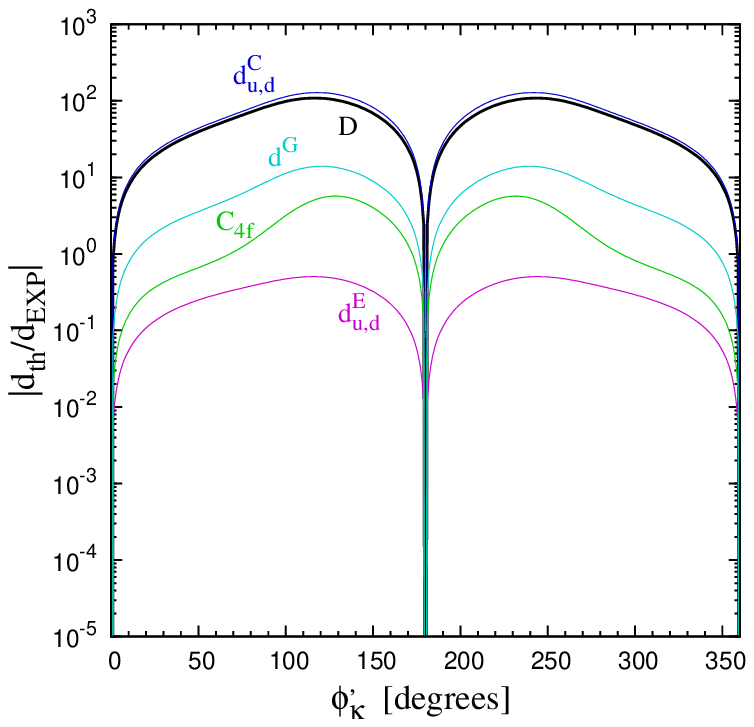}
\includegraphics[width=7.3cm]{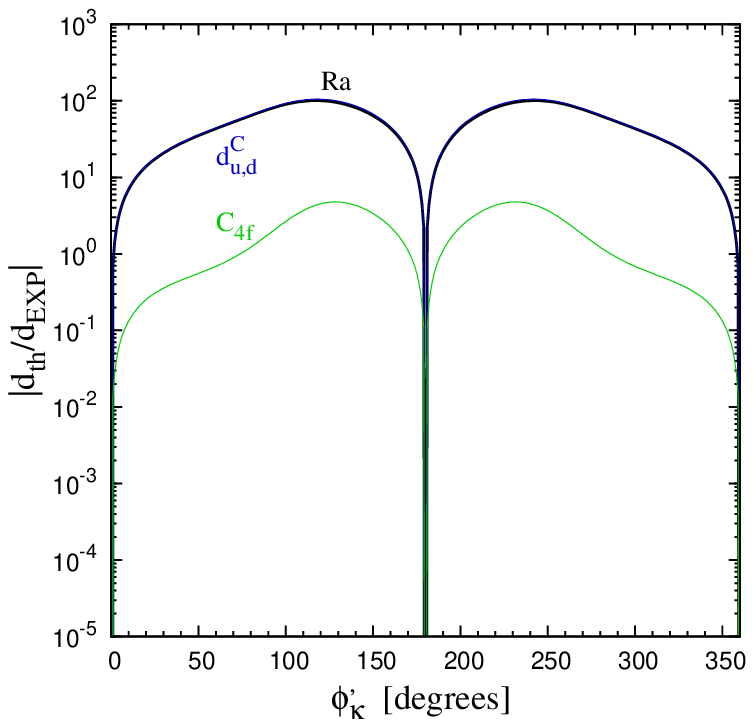}
\end{center}
\caption{The observable EDMs together with the
constituent contributions taking $|\lambda|=0.81$,
$|\kappa|=0.08$,
$|A_\lambda|=575$ GeV, and
$|A_\kappa|=110$ GeV.
The other parameters are fixed as in Eq.~(\ref{eq:scenario}).
The left (right) frame shows
the deuteron (Radium) EDM.}
\label{fig:S4_D_Ra}
\end{figure}
\begin{figure}[t!]
\begin{center}
\includegraphics[width=7.3cm]{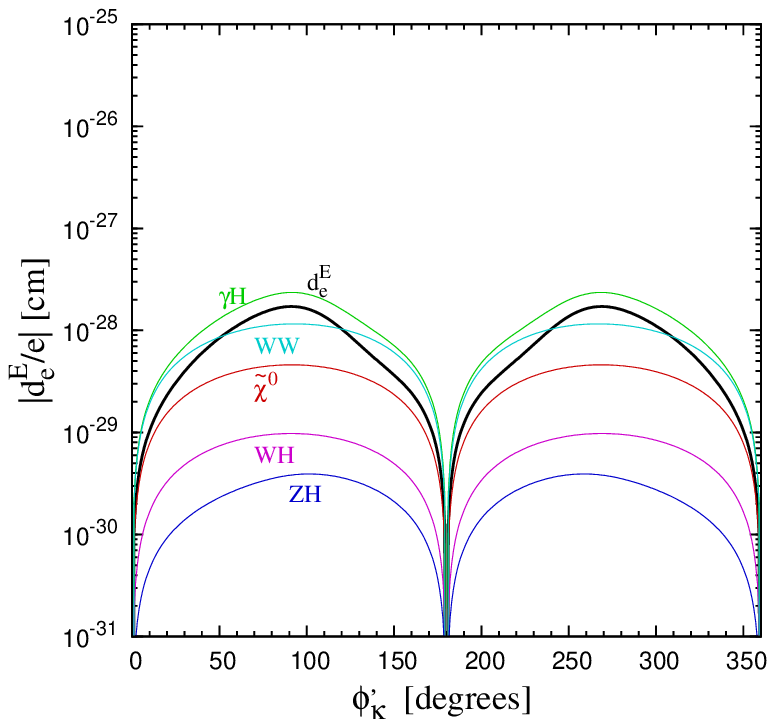}
\includegraphics[width=7.3cm]{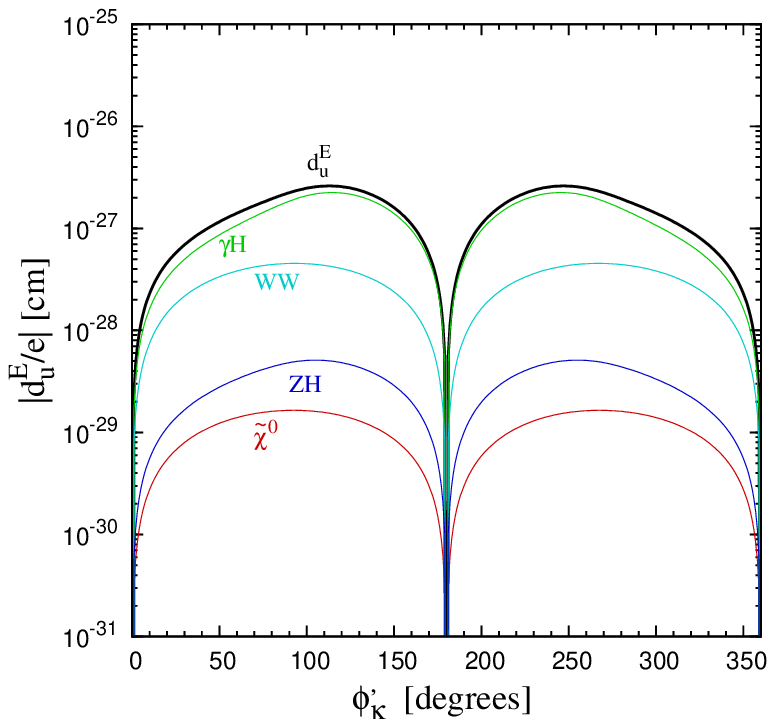}
\includegraphics[width=7.3cm]{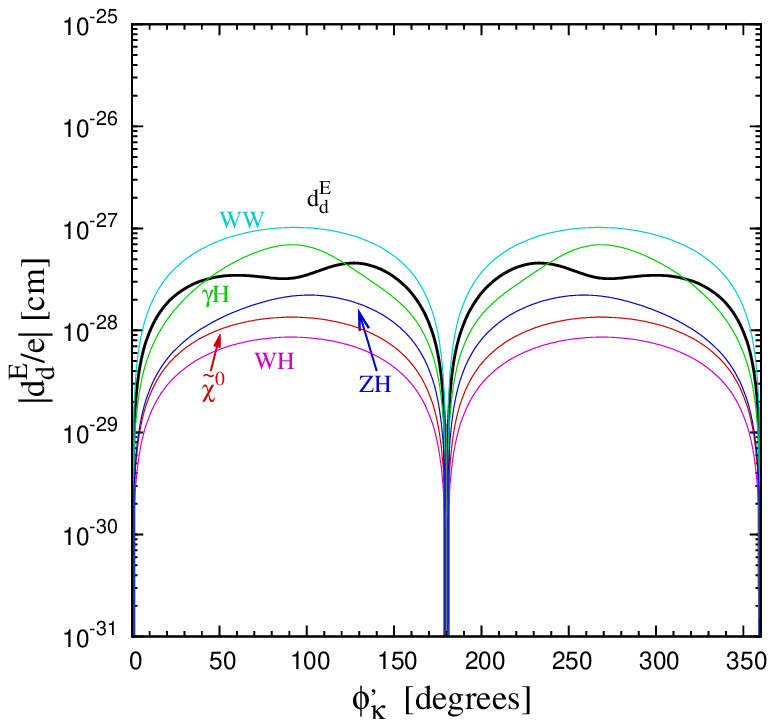}
\includegraphics[width=7.3cm]{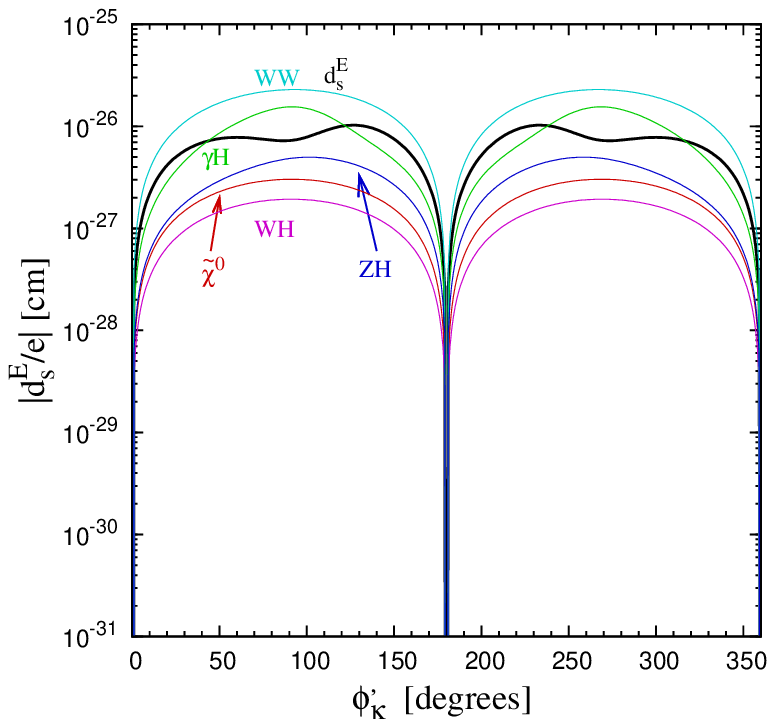}
\includegraphics[width=7.3cm]{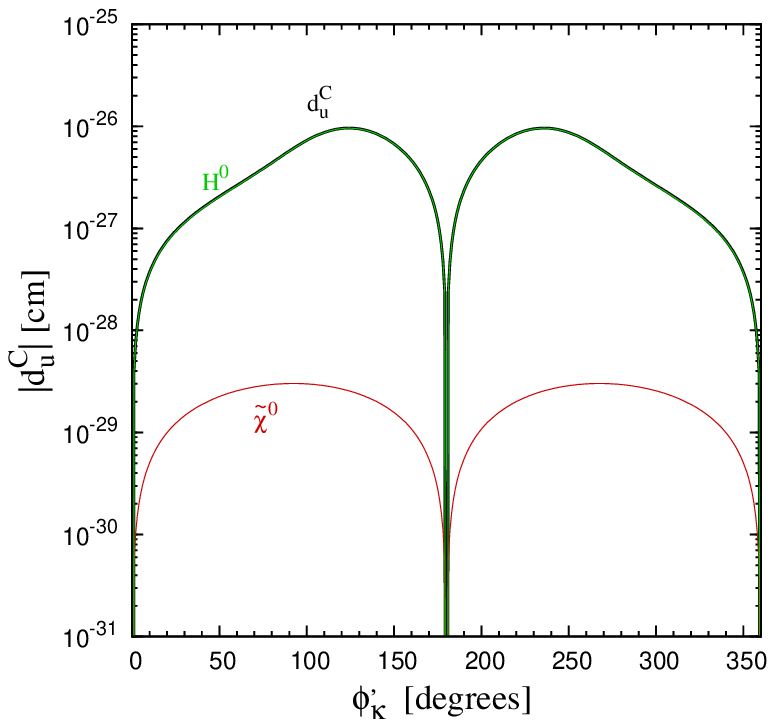}
\includegraphics[width=7.3cm]{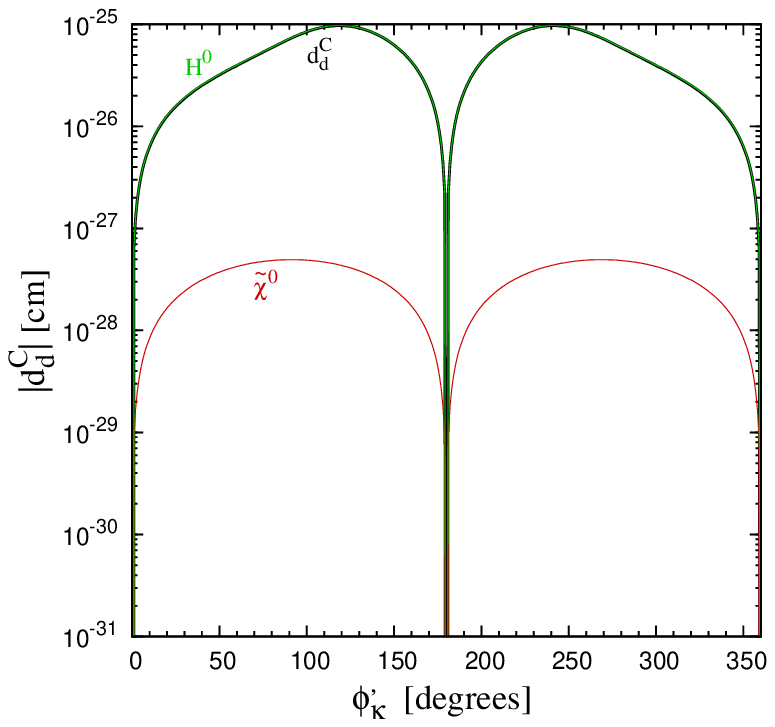}
\end{center}
\caption{The EDMs and CEDMs of the electron and 
the light quarks together with their constituent contributions
taking $|\lambda|=0.81$,
$|\kappa|=0.08$,
$|A_\lambda|=575$ GeV, and
$|A_\kappa|=110$ GeV.
The other parameters are fixed as in Eq.~(\ref{eq:scenario}).
Each frame shows
$|d^E_e/e|$ (upper left),
$|d^E_u/e|$ (upper right),
$|d^E_d/e|$ (middle left),
$|d^E_s/e|$ (middle right),
$|d^C_u|$ (lower left), and
$|d^C_d|$ (lower right) in units of cm.
}
\label{fig:S4_each}
\end{figure}

In the remaining part of this section, we wish to present the details
of the observable EDMs by exemplifying a point
\begin{equation}
|\lambda|=0.81\,, \\\
|\kappa|=0.08\,, \\\
|A_\lambda|=575~~{\rm GeV}\,,  \\\
|A_\kappa|=110~~{\rm GeV}.
\end{equation}
The other parameters are fixed as in Eq.~(\ref{eq:scenario}),
as motivated by EWBG.
In Fig.~\ref{fig:S4}, we show the absolute values of the
observable EDMs under consideration divided by 
the corresponding current experimental limits or 
the projected experimental sensitivities.

Fig.~\ref{fig:S4_Tl_n_Hg} shows the 
Thallium, neutron and Mercury EDMs together with the 
constituent contributions.
We observe that both the $d_e^E$ and $C_S$ terms significantly 
contribute to the Thallium EDM. The dominant
contributions to the neutron EDM in the QCD sum-rule approach
come from the CEDM and $d^G$ terms and we note that
the neutron EDM in the CQM shows the similar behavior (not shown).
The neutron EDM in the PQM is dominated by the contributions from
the EDMs of the up and strange quarks.
On the other hand,
the Mercury EDM is dominated by the CEDMs of the light quarks.
Fig.~\ref{fig:S4_D_Ra} shows the 
deuteron and Radium EDMs together with the 
constituent contributions. We observe both of the EDMs are dominated
by the contributions from the CEDM terms.

Fig.~\ref{fig:S4_each} shows the 
EDMs and CEDMS of the electron and light quarks together with the 
constituent contributions. 
We observe that the electron EDM is dominated
by the two-loop Barr--Zee contributions
mediated by the $\gamma$-$\gamma$-$H_i^0$  and
$\gamma$-$W^\pm$-$W^\mp$ couplings, whereas the one-loop 
contribution from the neutralino loops
is subleading. The one-loop contribution is suppressed 
because the CP phase $\phi_\kappa^\prime$ can contribute to EDM
only through the multiple singlino-Higgsino-gaugino mixing
and we are taking somewhat large values for the masses of 
the sfermions of the first two generations:
$M_{\widetilde{Q}_{1,2}}=M_{\widetilde{U}_{1,2}}=
M_{\widetilde{D}_{1,2}}=M_{\widetilde{L}_{1,2}}=M_{\widetilde{E}_{1,2}}=1$ TeV.
The other two-loop contributions mediated by the
$\gamma$-$H^\pm$-$W^\mp$ and $\gamma$-$H^0$-$Z$ couplings
are suppressed by the large charged Higgs-boson mass $\sim 600$ GeV
and the small vector coupling of the $Z$ boson to electrons,
$v_{Ze^+e^-}=-1/4+s_W^2$ with $s_W^2=0.23$, respectively.

The EDMs of the light quarks are dominated by the 
two-loop Barr--Zee contributions
mediated by the $\gamma$-$\gamma$-$H_i^0$  and
$\gamma$-$W^\pm$-$W^\mp$ couplings.
Being different from the case of $d^E_e$, the two-loop 
contribution mediated by the $\gamma$-$H^0$-$Z$ couplings
is larger than the one-loop contribution.
We note a cancellation occurs between the 
two dominant two-loop
Barr--Zee contributions
in $d^E_{d,s}$, resulting in 
$|d^E_{d}| < |d^E_{u}| \sim |d^E_{s}|$.
The CEDMs of the up and down quarks are
dominated by the Higgs-mediated two-loop contributions 
which are more than 100 times larger than the one-loop contributions
and we note $|d^C_{d}|/|d^C_{u}| \sim (m_d/m_u)\,\tan\beta \sim 10$.

\begin{figure}[t!]
\begin{center}
\includegraphics[width=17.3cm]{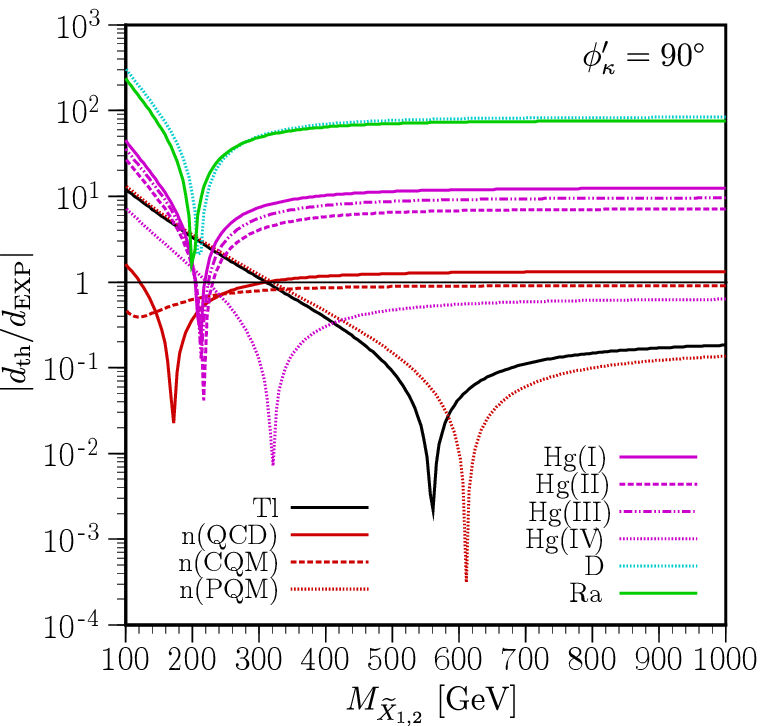}
\end{center}
\caption{The observable EDMs taking
$|\lambda|=0.81$,
$|\kappa|=0.08$,
$|A_\lambda|=575$ GeV, and
$|A_\kappa|=110$ GeV
as functions of 
$M_{\widetilde{X}_{1,2}}
\equiv M_{\widetilde{Q}_{1,2}}=M_{\widetilde{U}_{1,2}}=
M_{\widetilde{D}_{1,2}}=M_{\widetilde{L}_{1,2}}=M_{\widetilde{E}_{1,2}}$.
We have fixed $\phi_\kappa^\prime=90^\circ$ and
the other parameters are taken as in Eq.~(\ref{eq:scenario}).
The lines are the same as those in Fig.~\ref{fig:S4}.
}
\label{fig:S4_M12}
\end{figure}
%
Finally,
we examine the dependence of the observable EDMs on
the sfermion masses of the first two generations.
Fig.~\ref{fig:S4_M12} shows the observables EDMs as functions of
the common mass scale for the first two generations,
$M_{\widetilde{X}_{1,2}}
\equiv M_{\widetilde{Q}_{1,2}}=M_{\widetilde{U}_{1,2}}=
M_{\widetilde{D}_{1,2}}=M_{\widetilde{L}_{1,2}}=M_{\widetilde{E}_{1,2}}$.
Except the neutron EDM  based on the CQM which 
lies below the current experimental limit independently of
$M_{\widetilde{X}_{1,2}}$, all the EDMs exhibit dips
at certain values of $M_{\widetilde{X}_{1,2}}$. 
The dips occur because of
the cancellation between the one- and two-loop contributions.
When $M_{\widetilde{X}_{1,2}}$ is small the one-loop contribution dominates.
As $M_{\widetilde{X}_{1,2}}$ grows, the one-loop contribution
decouples and the EDMs saturate to certain values determined  
by the two-loop contribution.
Therefore, for the neutron (QCD), Mercury,
deuteron, and Radium EDMs,
we observe that the one-loop contribution is comparable to 
or larger than the two-loop one only
when $M_{\widetilde{X}_{1,2}} \lsim 300$ GeV.
On the other hand, for the neutron EDM based on the PQM and
the Thallium EDM, the one-loop contributions are larger but 
the two-loop contribution starts to dominate when
$M_{\widetilde{X}_{1,2}}$ is larger than $\sim 600$ GeV.
%
By choosing
$d^{\rm \,IV}_{\rm Hg}$ for the Mercury EDM,
we observe that all the EDM constraints
could be fulfilled when $M_{\widetilde{X}_{1,2}} \gsim 300$ GeV
without relying on the cancellation mechanism.
If we make other choices for the Mercury EDM,  to suppress all the
EDMs for Thallium, neutron, and Mercury below their
present experimental bounds,
the required degree of cancellation is about
90 \% over the whole range of $M_{\widetilde{X}_{1,2}}$,
with  100 \%  corresponding to complete cancellation.

Before we close this section,
we make a comment on the $\tan\beta$ dependence of
the EDMs of the electron and the down and strange quarks.
The one-loop neutralino contribution is
proportional to $\tan\beta$.
The Barr--Zee contribution mediated by the $\gamma$-$W^\pm$-$W^\mp$ couplings,
one of the two two-loop leading contributions,
is nearly independent of $\tan\beta$. The $\tan\beta$ dependence of
the other dominant Barr--Zee contribution
mediated by the $\gamma$-$\gamma$-$H_i^0$ couplings is
much milder compared to the MSSM case in which
the sbottom and stau contributions are proportional 
to $\tan^3\beta$~\cite{Chang:1998uc,More_twoloops,Demir:2003js}.
This is because the masses
of the two heavy Higgs states, which include the CP-odd state from
the Higgs doublets, increase as $\tan\beta$ grows,
$M_{H_4,H_5} \simeq (|\lambda|v_S/\sqrt{2})\,\tan\beta$, to
avoid tachyonic Higgs states~\cite{Cheung:2010ba}.

\section{Conclusions}

We have performed a study on the predictions of EDMs for
Thallium, neutron, Mercury, deuteron, and Radium in the framework of
CP violating NMSSM.
The rephasing invariant combinations of the physical CP phases
contributing to the EDMs are 
$(\phi_\lambda^\prime+\phi_{A_f})$,
$(\phi_\lambda^\prime+\phi_i)$, and
$(\phi_\lambda^\prime-\phi_\kappa^\prime)$,
with $\phi_{A_f}$ and $\phi_i$ denoting the CP phases of the soft 
trilinear parameters $A_f$ and the three gaugino mass 
parameters $M_{i=1,2,3}$, respectively.
Unlike the MSSM the non-vanishing
CP phase $(\phi_\lambda^\prime - \phi_\kappa^\prime)$ could 
result in significant
CP-violating mixing among the neutral Higgs bosons 
at tree level even when $\sin(\phi_\lambda^\prime+\phi_{A_f})
=\sin(\phi_\lambda^\prime+\phi_i)=0$.
Throughout this work, we have taken a convention of $\phi_\lambda^\prime=0$.

The MSSM CP phases $\phi_{A_f}$ and $\phi_i$ are generally known to be
strongly constrained by the non-observation of 
the Thallium, neutron, and Mercury EDMs, without invoking 
cancellations among various comparable contributions, if the sfermions
are within the reach of the LHC. On the other hand,
the matter-dominated Universe requires
sources of CP violation other than 
the single Kobayashi--Maskawa phase
\footnote{See Ref.~\cite{Li:2008ez} for the recently suggested
bino-driven EWBG scenario exploiting the CP phase of the bino mass parameter,
$\phi_1$, to account for successful EWBG without
inducing large EDMs.}.
In this work, we concentrate on the CP phase
$\phi_\kappa^\prime$, which only exist
in the NMSSM framework, by taking $\sin(\phi_{A_f})=\sin(\phi_i)=0$.

One of the attractive features of the NMSSM,
compared to the MSSM, might be that the mechanism of 
EWBG could be realized in a more natural setting.
In analyzing the EDM constraint on the CP phase
$\phi_\kappa^\prime$, we have taken a scenario 
in which a first-order phase transition is presumed to occur.
The strong enough first-order phase transition
is essential to the EWBG. 

Previously, the constraint on $\phi_\kappa^\prime$ 
from the neutron and electron (equivalently, Thallium) EDMs
had been considered but only the 
one-loop neutralino contributions were taken into account. 
We check that our one-loop results agree well 
with the previous ones, showing no strong
constraints on $\phi_\kappa^\prime$ 
especially when the sfermions of the first two generations
are heavier than $\sim$ 300 GeV.  
In this work, we have further considered the constraint from
the non-observation of Mercury EDM and found the similar results.

In addition to the one-loop contributions, we have taken account of
the higher-order corrections to the EDMs and CEDMs of the light quarks
and electron, the dimension-six Weinberg 
and the Higgs-exchange four-fermion operators.
We note that most of these operators are generated due to the 
CP-violating neutral Higgs-boson mixing induced by the CP phase
$\phi_\kappa^\prime$.
The two-loop contributions 
and, especially, the coefficient of  the dimension-six
Weinberg operator start to dominate
when the sfermions of the first two generations
are heavier than $\sim$ 300 GeV.
We found that they can saturate the current bound 
on the neutron EDM and can go over that on 
the Mercury EDM.
For the Mercury EDM,
we have included the uncertainties in the calculation of
the Schiff-moment-induced term and
found that there is still a room to have the maximal
CP phase $\phi_\kappa^\prime \sim 90^\circ$.
We have also shown that the large CP phase 
$\phi_\kappa^\prime \sim 90^\circ$
can be easily probed in
the proposed future experiments searching for the EDMs of
deuteron and the $^{225}$Ra atom and it might be
connected to the EWBG, providing a new mechanism for it~\cite{future1}.

Furthermore, we offer a few more comments as follows before we close.
\begin{enumerate}
\item The new CP phase that we considered here can be applied to
a number of low-energy phenomenologies, such as CP asymmetries in $B$
mesons and $K$ mesons.

\item In our previous work \cite{Cheung:2010ba}, we have imposed 
the following constraints to the parameter space:
the LEP limits, the global minimum condition, and the positivity of the 
square of the Higgs-boson masses. In the present work, we have 
further constrained the parameter space required by the EDM constraints. 
Therefore, by combining all these constraints we can explore further
phenomenologies, including the Higgs physics at the LHC, the muon
anomalous magnetic moment, and electroweak baryogenesis.

\item One of the reasons why the 2-loop BZ diagrams can dominate is 
that one or more of the neutral Higgs bosons become very light.  
One can imagine that contributions to the muon anomalous magnetic moments
can also become important in comparison to the one-loop result. In fact, one
can show that the muon anomalous magnetic moment and EDM are related
to the real and imaginary parts of the combination of couplings.
We will come back to this issue in a future work.

\item We will soon make available a computer code for calculating 
all the couplings and masses of the Higgs bosons, with the parameter
space restricted by all the experimental constraints (all the above
mentioned ones as well as direct search limits, muon $g-2$, etc) as
we proceed further in this framework.

\end{enumerate}
%

\vspace{-0.2cm}
\subsection*{Acknowledgements}
\vspace{-0.3cm}
\noindent
The work was supported in parts
by the NSC of Taiwan (grant No. 99-2112-M-007-005-MY3), the NCTS, 
and by the WCU program through the KOSEF funded by
the MEST (R31-2008-000-10057-0).

%
\section*{Appendices}

\def\theequation{\Alph{section}.\arabic{equation}}
\begin{appendix}
Throughout Appendices, we are following the conventions and notations of {\tt
CPsuperH}~\cite{cpsuperh}.
\setcounter{equation}{0}
\section{Masses and mixing matrices}
Here we set up our conventions and notations of the masses and mixing matrices 
of neutral Higgs bosons, charginos, neutralinos, third-generations sfermions.
\begin{itemize}
\item \underline{Neutral Higgs bosons}:
\begin{equation}
(\phi_d^0,\phi_u^0,\phi_S^0,a,a_S)^T_\alpha
=O_{\alpha i}(H_1,H_2,H_3,H_4,H_5)^T_i \,,
\end{equation}
where $O^T{\cal M}^2_H\, O={\sf diag} \, (M^2_{H_1}, M^2_{H_2},
M^2_{H_3},M^2_{H_4}, M^2_{H_5})$ with
$M_{H_1} \leq M_{H_2}\leq M_{H_3} \leq M_{H_4}\leq M_{H_5}$.
\item \underline{Charginos}:
We adopt the convention
$\wt{H}^{-}_{L(R)} = \wt{H}^{-}_{d(u)}$.
The chargino mass matrix in the $(\widetilde{W}^-,\widetilde{H}^-)$ basis
\begin{eqnarray}
{\cal M}_C = \left(\begin{array}{cc}
     M_2              & \sqrt{2} M_W\, c_{\beta} \\[2mm]
\sqrt{2} M_W\, s_{\beta} &
\frac{|\lambda| v_S}{\sqrt{2}}\,e^{i(\phi_\lambda+\theta+\varphi)}
             \end{array}\right)\, ,
\end{eqnarray}
is diagonalized by two different unitary matrices
$ C_R{\cal M}_C C_L^\dagger ={\sf diag}\{m_{\widetilde{\chi}^\pm_1},\,
m_{\widetilde{\chi}^\pm_2}\}$, where
$m_{\widetilde{\chi}^\pm_1} \leq m_{\widetilde{\chi}^\pm_2}$.
The chargino mixing matrices $(C_L)_{i\alpha}$ and $(C_R)_{i\alpha}$
relate the electroweak eigenstates to the mass eigenstates, via
\begin{eqnarray}
\widetilde{\chi}^-_{\alpha L} &=&
(C_L)^*_{i \alpha } \widetilde{\chi}_{iL}^-\,,\qquad
\widetilde{\chi}^-_{\alpha L}\ =\ (\widetilde{W}^-, \widetilde{H}^-)_L^T\,,\nonumber\\
\widetilde{\chi}^-_{\alpha R} &=& (C_R)^*_{i \alpha} \widetilde{\chi}_{iR}^-\,,\qquad
\widetilde{\chi}^-_{\alpha R}\ =\ (\widetilde{W}^-, \widetilde{H}^-)_R^T\,.
\end{eqnarray}
We use the following abbreviations throughout this paper:
$s_\beta\equiv\sin\beta$, $c_\beta\equiv\cos\beta$,
$t_\beta=\tan\beta$, $s_{2\beta}\equiv\sin\,2\beta$,
$c_{2\beta}\equiv\cos\,2\beta$, $s_W\equiv\sin\theta_W$,
$c_W\equiv\cos\theta_W$, etc.
\item \underline{Neutralinos}:
The symmetric neutralino mass matrix in the
$(\wt{B},\wt{W}^0,\wt{H}^0_d,\wt{H}^0_u,\wt{S})$ basis is given by
\begin{eqnarray}
{\cal M}_N=\left(\begin{array}{ccccc}
  M_1       &      0          &  -M_Z c_\beta s_W  & M_Z s_\beta s_W  & 0\\[2mm]
 &     M_2         &   M_Z c_\beta c_W  & -M_Z s_\beta c_W & 0\\[2mm]
  & & 0 &
-\frac{|\lambda| v_S}{\sqrt{2}}\,e^{i(\phi_\lambda+\theta+\varphi)} &
-\frac{|\lambda| v s_\beta}{\sqrt{2}}\,e^{i(\phi_\lambda+\theta+\varphi)} \\[2mm]
 & &  &       0          &
-\frac{|\lambda| v c_\beta}{\sqrt{2}}\,e^{i(\phi_\lambda+\theta+\varphi)} \\[2mm]
& & & &
\sqrt{2} |\kappa| v_S \,e^{i(\phi_\kappa+3\varphi)}
                  \end{array}\right)\,
\end{eqnarray}
This neutralino mass matrix is diagonalized by a unitary matrix $N$:
$N^* {\cal M}_N N^\dagger = {\sf diag}\,
(m_{\widetilde{\chi}_1^0},m_{\widetilde{\chi}_2^0},
m_{\widetilde{\chi}_3^0},m_{\widetilde{\chi}_4^0},m_{\widetilde{\chi}_5^0})$
with
$m_{\widetilde{\chi}_1^0} \leq m_{\widetilde{\chi}_2^0} \leq m_{\widetilde{\chi}_3^0}
\leq m_{\widetilde{\chi}_4^0} \leq m_{\widetilde{\chi}_5^0}$.
The neutralino mixing matrix $N_{i\alpha}$
relates the electroweak eigenstates to the mass eigenstates via
\begin{eqnarray}
(\wt{B},\wt{W}^0,\wt{H}^0_d,\wt{H}^0_u,\wt{S})^T_{\alpha}
=N_{i\alpha}^*
({\widetilde{\chi}_1^0},{\widetilde{\chi}_2^0},{\widetilde{\chi}_3^0},
{\widetilde{\chi}_4^0},{\widetilde{\chi}_5^0})^T_{i}\,.
\end{eqnarray}
%
\item \underline{Stops, sbottoms, staus and tau sneutrino}:
At the tree level, the Yukawa couplings are given by
\begin{equation}
h_l = \frac{\sqrt{2}\,m_l}{ v c_\beta}\,; \ \ \
h_d = \frac{\sqrt{2}\,m_d}{ v c_\beta}\,; \ \ \
h_u = e^{-i\theta}\,\frac{\sqrt{2}\,m_u}{ v s_\beta}\,.
\end{equation}
The stop and sbottom mass
matrices may conveniently be written in the $\left(\widetilde{q}_L,
\widetilde{q}_R\right)$ basis as
\begin{eqnarray}
  \label{Msuark}
\hspace{-0.5 cm}
\widetilde{\cal M}^2_t  & = & \left( \begin{array}{cc}
M^2_{\widetilde{Q}_3}\, +\, m^2_t\, +\, c_{2\beta} M^2_Z\, ( T^t_{3L}\, -\,
Q_t s_W^2 ) &
h_t^* v_u (A^*_u e^{-i\theta} -
\frac{|\lambda| v_S}{\sqrt{2}}e^{i(\phi_\lambda+\varphi)} \cot\beta )/\sqrt{2}\\
h_t v_u (A_u e^{i\theta} -
\frac{|\lambda| v_S}{\sqrt{2}}e^{-i(\phi_\lambda+\varphi)} \cot\beta )/\sqrt{2}
& \hspace{-0.2cm}
M^2_{\widetilde{U}_3}\, +\, m^2_t\, +\, c_{2\beta} M^2_Z\, Q_t s^2_W
\end{array}\right)\, ,
\nonumber \\[4mm]
\hspace{-0.5 cm}
\widetilde{\cal M}^2_b  & = & \left( \begin{array}{cc}
M^2_{\widetilde{Q}_3}\, +\, m^2_b\, +\, c_{2\beta} M^2_Z\, ( T^b_{3L}\, -\,
Q_b s_W^2 ) &
h_b^* v_d (A^*_b -
\frac{|\lambda| v_S}{\sqrt{2}}e^{i(\phi_\lambda+\theta+\varphi)} \tan\beta )/\sqrt{2}\\
h_b v_d (A_b -
\frac{|\lambda| v_S}{\sqrt{2}}e^{-i(\phi_\lambda+\theta+\varphi)} \tan\beta )/\sqrt{2}
& \hspace{-0.2cm}
M^2_{\widetilde{D}_3}\, +\, m^2_b\, +\, c_{2\beta} M^2_Z\, Q_b s^2_W
\end{array}\right)\,, \nonumber \\
\end{eqnarray}
with
$T^t_{3L} = - T^b_{3L} = 1/2$,
$Q_t = 2/3$, $Q_b = -1/3$, and
$h_q$ is the Yukawa coupling of the quark $q$.
On the other hand, the stau mass matrix is written in the $\left(\widetilde{\tau}_L,
\widetilde{\tau}_R\right)$ basis as
\begin{equation}
  \label{Mstau}
\hspace{-0.5 cm}
\widetilde{\cal M}^2_\tau  = \left( \begin{array}{cc}
M^2_{\widetilde{L}_3}\, +\, m^2_\tau\, +\, c_{2\beta} M^2_Z\,
( s_W^2-1/2 ) &
h_\tau^* v_d (A^*_\tau -
\frac{|\lambda| v_S}{\sqrt{2}}e^{i(\phi_\lambda+\theta+\varphi)} \tan\beta )/\sqrt{2}\\
h_\tau v_d (A_\tau -
\frac{|\lambda| v_S}{\sqrt{2}}e^{-i(\phi_\lambda+\theta+\varphi)} \tan\beta )/\sqrt{2}
& \hspace{-0.2cm}
M^2_{\widetilde{E}_3}\, +\, m^2_\tau\, -\, c_{2\beta} M^2_Z\, s^2_W
\end{array}\right)\, ,
\end{equation}
and the mass of the tau sneutrino $\widetilde{\nu_\tau}$ is simply
$m_{\widetilde\nu_\tau} = \sqrt{ M^2_{\widetilde{L}_3} + \frac{1}{2}
c_{2\beta} M_Z^2 }$, as it has no right--handed counterpart in the
NMSSM as in the MSSM.

\medskip

The $2\times 2$ sfermion mass matrix $\widetilde{\cal M}^2_f$
for $f=t, b$ and $\tau$ is diagonalized by a unitary matrix
$U^{\widetilde{f}}$: $U^{\widetilde{f}\dagger} \, \widetilde{\cal M}^2_f \,
U^{\widetilde{f}} ={\sf diag}(m_{\widetilde{f}_1}^2,m_{\widetilde{f}_2}^2)\,$ with
$m_{\widetilde{f}_1}^2 \leq m_{\widetilde{f}_2}^2$.  The mixing matrix
$U^{\widetilde{f}}$ relates the electroweak eigenstates $\widetilde{f}_{L,R}$
to the mass eigenstates $\widetilde{f}_{1,2}$, via
\begin{equation}
(\widetilde{f}_L,\widetilde{f}_R)^T_\alpha\,=\,
U^{\widetilde{f}}_{\alpha i} \,
(\widetilde{f}_1,\widetilde{f}_2)^T_i\,.
\end{equation}
%
\end{itemize}
\setcounter{equation}{0}
\section{Higgs-boson couplings to sfermions}
Here we present the couplings of the neutral and charged Higgs bosons to 
squarks and sleptons in the weak basis.
\begin{itemize}
\item \underline{The neutral Higgs couplings to sfermions}:
\begin{itemize}
\item
In the $(\widetilde{b}_L,\widetilde{b}_R)$ basis, the neutral Higgs couplings to
the sbottoms:
\begin{eqnarray}
%
%
\Gamma^{\phi_d^0\,\widetilde{b}^*\widetilde{b}} &=& \left(
\begin{array}{cc}
-|h_b|^2vc_\beta+ \frac{1}{4}\left(g^2+\frac{1}{3}g^{\prime 2}\right)vc_\beta&
-\frac{1}{\sqrt{2}}h_b^*A_b^* \\[2mm]
-\frac{1}{\sqrt{2}}h_bA_b &
-|h_b|^2vc_\beta+ \frac{1}{6}g^{\prime2} vc_\beta
\end{array} \right)\,,
\nonumber \\[4mm]
\Gamma^{\phi_u^0\widetilde{b}^*\widetilde{b}} &=& \left(
\begin{array}{cc}
- \frac{1}{4}\left(g^2+\frac{1}{3}g^{\prime 2}\right)vs_\beta&
\frac{1}{2}h_b^*|\lambda | v_S \,e^{i(\phi_\lambda+\theta+\varphi)} \\[2mm]
\frac{1}{2}h_b|\lambda | v_S \,e^{-i(\phi_\lambda+\theta+\varphi)} &
-\frac{1}{6}g^{\prime2} vs_\beta
\end{array} \right)\,,
\nonumber \\[4mm]
\Gamma^{\phi_S^0\,\widetilde{b}^*\widetilde{b}} &=& \left(
\begin{array}{cc}
0 & \frac{1}{2}\,h_b^*
|\lambda |\, v\,s_\beta\,
e^{i(\phi_\lambda + \theta +\varphi)} \\[2mm]
\frac{1}{2}\,h_b
|\lambda |\, v\,s_\beta\,
e^{-i(\phi_\lambda + \theta +\varphi)}
& 0
\end{array} \right)\,,
\nonumber \\[4mm]
\Gamma^{a\,\widetilde{b}^*\widetilde{b}} &=& \frac{1}{\sqrt{2}}\left(
\begin{array}{cc}
0 & i\,h_b^*\left(s_\beta A_b^*+c_\beta
\frac{|\lambda |\, v_S}{\sqrt{2}}\,
e^{i(\phi_\lambda + \theta +\varphi)}\right) \\[2mm]
-i\,h_b\left(s_\beta A_b+c_\beta
\frac{|\lambda |\, v_S}{\sqrt{2}}\,
e^{-i(\phi_\lambda + \theta +\varphi)}
\right) & 0
\end{array} \right)\,,
\nonumber \\[4mm]
\Gamma^{a_S\,\widetilde{b}^*\widetilde{b}} &=& \left(
\begin{array}{cc}
0 & i\,\frac{1}{2}\,h_b^*
|\lambda |\, v\,s_\beta\,
e^{i(\phi_\lambda + \theta +\varphi)} \\[2mm]
-i\,\frac{1}{2}\,h_b
|\lambda |\, v\,s_\beta\,
e^{-i(\phi_\lambda + \theta +\varphi)}
& 0
\end{array} \right)\,, \nonumber 
\end{eqnarray}
\item
In the $(\widetilde{t}_L,\widetilde{t}_R)$ basis, the neutral Higgs couplings to
the stops:
\begin{eqnarray}
%
%
\Gamma^{\phi_d^0\,\widetilde{t}^*\widetilde{t}} &=& \left(
\begin{array}{cc}
- \frac{1}{4}\left(g^2-\frac{1}{3}g^{\prime 2}\right)vc_\beta &
\frac{1}{2}h_t^*|\lambda | v_S\,e^{i(\phi_\lambda+\varphi)} \\[2mm]
\frac{1}{2}h_t|\lambda | v_S\,e^{-i(\phi_\lambda+\varphi)} &
-\frac{1}{3}g^{\prime2} vc_\beta
\end{array} \right)\,,
\nonumber \\[4mm]
\Gamma^{\phi_u^0\,\widetilde{t}^*\widetilde{t}} &=& \left(
\begin{array}{cc}
-|h_t|^2vs_\beta+ \frac{1}{4}\left(g^2-\frac{1}{3}g^{\prime 2}\right)vs_\beta&
-\frac{1}{\sqrt{2}}h_t^*A_t^*\,e^{-i\theta} \\[2mm]
-\frac{1}{\sqrt{2}}h_tA_t\,e^{i\theta} &
-|h_t|^2vs_\beta+ \frac{1}{3}g^{\prime2} vs_\beta
\end{array} \right)\,,
\nonumber \\[4mm]
\Gamma^{\phi_S^0\,\widetilde{t}^*\widetilde{t}} &=& \left(
\begin{array}{cc}
0 & \frac{1}{2}\,h_t^*
|\lambda |\, v\,c_\beta\,
e^{i(\phi_\lambda + \varphi)} \\[2mm]
\frac{1}{2}\,h_t
|\lambda |\, v\,c_\beta\,
e^{-i(\phi_\lambda +\varphi)}
& 0
\end{array} \right)\,,
\nonumber \\[4mm]
\Gamma^{a\,\widetilde{t}^*\widetilde{t}} &=& \frac{1}{\sqrt{2}}\left(
\begin{array}{cc}
0 & i\,h_t^*\left(c_\beta A_t^*\,e^{-i\theta}+
s_\beta \frac{|\lambda |\,v_S}{\sqrt{2}}\,e^{i(\phi_\lambda+\varphi)}
\right) \\[2mm]
-i\,h_t\left(c_\beta A_t\,e^{i\theta}+
s_\beta \frac{|\lambda |\,v_S}{\sqrt{2}}\,e^{-i(\phi_\lambda+\varphi)}
\right) & 0
\end{array} \right)\,,
\nonumber \\[4mm]
\Gamma^{a_S\,\widetilde{t}^*\widetilde{t}} &=& \left(
\begin{array}{cc}
0 & i\,\frac{1}{2}\,h_t^*
|\lambda |\, v\,c_\beta\,
e^{i(\phi_\lambda + \varphi)} \\[2mm]
-i\,\frac{1}{2}\,h_t
|\lambda |\, v\,c_\beta \,
e^{-i(\phi_\lambda +\varphi)}
& 0
\end{array} \right)\,, \nonumber 
\end{eqnarray}
\item
In the $(\widetilde{\tau}_L,\widetilde{\tau}_R)$ basis, the neutral Higgs couplings to
the staus:
\begin{eqnarray}
%
%
\Gamma^{\phi_d^0\,\widetilde{\tau}^*\widetilde{\tau}} &=& \left(
\begin{array}{cc}
-|h_\tau|^2vc_\beta+ \frac{1}{4}\left(g^2-g^{\prime 2}\right)vc_\beta&
-\frac{1}{\sqrt{2}}h_\tau^*A_\tau^* \\[2mm]
-\frac{1}{\sqrt{2}}h_\tau A_\tau &
-|h_\tau|^2vc_\beta+ \frac{1}{2}g^{\prime2} vc_\beta
\end{array} \right)\,,
\nonumber \\[4mm]
\Gamma^{\phi_u^0\widetilde{\tau}^*\widetilde{\tau}} &=& \left(
\begin{array}{cc}
- \frac{1}{4}\left(g^2-g^{\prime 2}\right)vs_\beta&
\frac{1}{2}h_\tau^*|\lambda | v_S \,e^{i(\phi_\lambda+\theta+\varphi)} \\[2mm]
\frac{1}{2}h_\tau|\lambda | v_S \,e^{-i(\phi_\lambda+\theta+\varphi)} &
-\frac{1}{2}g^{\prime2} vs_\beta
\end{array} \right)\,,
\nonumber \\[4mm]
\Gamma^{\phi_S^0\,\widetilde{\tau}^*\widetilde{\tau}} &=& \left(
\begin{array}{cc}
0 & \frac{1}{2}\,h_\tau^*
|\lambda |\, v\,s_\beta\,
e^{i(\phi_\lambda + \theta +\varphi)} \\[2mm]
\frac{1}{2}\,h_\tau
|\lambda |\, v\,s_\beta\,
e^{-i(\phi_\lambda + \theta +\varphi)}
& 0
\end{array} \right)\,,
\nonumber \\[4mm]
\Gamma^{a\,\widetilde{\tau}^*\widetilde{\tau}} &=& \frac{1}{\sqrt{2}}\left(
\begin{array}{cc}
0 & i\,h_\tau^*\left(s_\beta A_\tau^*+c_\beta
\frac{|\lambda |\, v_S}{\sqrt{2}}\,
e^{i(\phi_\lambda + \theta +\varphi)}\right) \\[2mm]
-i\,h_\tau\left(s_\beta A_\tau+c_\beta
\frac{|\lambda |\, v_S}{\sqrt{2}}\,
e^{-i(\phi_\lambda + \theta +\varphi)}
\right) & 0
\end{array} \right)\,,
\nonumber \\[4mm]
\Gamma^{a_S\,\widetilde{\tau}^*\widetilde{\tau}} &=& \left(
\begin{array}{cc}
0 & i\,\frac{1}{2}\,h_\tau^*
|\lambda |\, v\,s_\beta\,
e^{i(\phi_\lambda + \theta +\varphi)} \\[2mm]
-i\,\frac{1}{2}\,h_\tau
|\lambda |\, v\,s_\beta\,
e^{-i(\phi_\lambda + \theta +\varphi)}
& 0
\end{array} \right)\,, \nonumber
\end{eqnarray}
\item The neutral Higgs couplings to the sneutrinos
\begin{eqnarray}
\Gamma^{\phi_1\widetilde{\nu}^*_\tau\widetilde{\nu}_\tau} = -\frac{1}{4}
\left( g^2 + g^{\prime 2} \right) v c_\beta,
\qquad
\Gamma^{\phi_2\widetilde{\nu}^*_\tau\widetilde{\nu}_\tau} = \frac{1}{4} \left(
g^2 + g^{\prime 2} \right) v s_\beta \, , \nonumber
\end{eqnarray}
and the other couplings are vanishing.
\end{itemize}
\item \underline{The charged Higgs couplings to sfermions}:
\begin{eqnarray}
\Gamma^{H^+\widetilde{u}^*\widetilde{d}}\ &=&\ \left(
\begin{array}{cc} \frac{1}{\sqrt{2}}\,(|h_u|^2 + |h_d|^2
- g^2)\,v s_\beta c_\beta &
h_d^*\, \left( s_\beta A^*_d + c_\beta
\frac{|\lambda |\,v_S}{\sqrt{2}}\,e^{i(\phi_\lambda+\theta+\varphi)} \right)\\[2mm]
h_u\,\left( c_\beta A_u\,e^{i\theta} + s_\beta
\frac{|\lambda |\,v_S}{\sqrt{2}}\,e^{-i(\phi_\lambda+\varphi)} \right) &
\frac{1}{\sqrt{2}}\, h_u h^*_d\, v \end{array}\right)\,
\nonumber\\[4mm]
\Gamma^{H^+\widetilde{\nu}_\tau^*\widetilde{\tau}_L}\ &=& \
 \frac{1}{\sqrt{2}}\,(|h_\tau|^2 - g^2)\,v s_\beta\, c_\beta\,,\qquad
\Gamma^{H^+\widetilde{\nu}_\tau^*\widetilde{\tau}_R}\ = \
  h^*_\tau \left(s_\beta A^*_\tau+c_\beta
\frac{|\lambda |\,v_S}{\sqrt{2}}\,e^{i(\phi_\lambda+\theta+\varphi)} \right)\,.
\nonumber 
\end{eqnarray}
\end{itemize}

\end{appendix}

%


\begin{thebibliography}{99}

\bibitem{NMSSM:0}
  M.~Dine, W.~Fischler and M.~Srednicki,
  Phys.\ Lett.\  B {\bf 104} (1981) 199;
  H.~P.~Nilles, M.~Srednicki and D.~Wyler,
  Phys.\ Lett.\  B {\bf 120} (1983) 346;
  J.~M.~Frere, D.~R.~T.~Jones and S.~Raby,
  Nucl.\ Phys.\  B {\bf 222} (1983) 11;
  A.~I.~Veselov, M.~I.~Vysotsky and K.~A.~Ter-Martirosian,
  Sov.\ Phys.\ JETP {\bf 63} (1986) 489
  [Zh.\ Eksp.\ Teor.\ Fiz.\  {\bf 90} (1986) 838];
  J.~P.~Derendinger and C.~A.~Savoy,
  Nucl.\ Phys.\  B {\bf 237} (1984) 307.

\bibitem{NMSSM:1}
  J.~R.~Ellis, J.~F.~Gunion, H.~E.~Haber, L.~Roszkowski and F.~Zwirner,
  Phys.\ Rev.\  D {\bf 39} (1989) 844;
  U.~Ellwanger,
  Phys.\ Lett.\  B {\bf 303} (1993) 271
  [arXiv:hep-ph/9302224];
  U.~Ellwanger, M.~Rausch de Traubenberg and C.~A.~Savoy,
  Phys.\ Lett.\  B {\bf 315} (1993) 331
  [arXiv:hep-ph/9307322];
  P.~N.~Pandita,
  Phys.\ Lett.\  B {\bf 318} (1993) 338;
  P.~N.~Pandita,
  Z.\ Phys.\  C {\bf 59} (1993) 575;
  T.~Elliott, S.~F.~King and P.~L.~White,
  Phys.\ Rev.\  D {\bf 49} (1994) 2435
  [arXiv:hep-ph/9308309];
  G.~K.~Yeghian,
  arXiv:hep-ph/9904488;
  S.~F.~King and P.~L.~White,
  Phys.\ Rev.\  D {\bf 52} (1995) 4183
  [arXiv:hep-ph/9505326];
  U.~Ellwanger, M.~Rausch de Traubenberg and C.~A.~Savoy,
  Nucl.\ Phys.\  B {\bf 492} (1997) 21
  [arXiv:hep-ph/9611251];
  F.~Franke and H.~Fraas,
  Int.\ J.\ Mod.\ Phys.\  A {\bf 12} (1997) 479
  [arXiv:hep-ph/9512366];
  B.~Ananthanarayan and P.~N.~Pandita,
  Int.\ J.\ Mod.\ Phys.\  A {\bf 12} (1997) 2321
  [arXiv:hep-ph/9601372];
  U.~Ellwanger and C.~Hugonie,
  Eur.\ Phys.\ J.\  C {\bf 25} (2002) 297
  [arXiv:hep-ph/9909260];
  U.~Ellwanger, J.~F.~Gunion, C.~Hugonie and S.~Moretti,
  arXiv:hep-ph/0305109;
  U.~Ellwanger and C.~Hugonie,
  Phys.\ Lett.\  B {\bf 623} (2005) 93
  [arXiv:hep-ph/0504269].

\bibitem{NMSSM:ECPV}
  M.~Matsuda and M.~Tanimoto,
  Phys.\ Rev.\  D {\bf 52} (1995) 3100
  [arXiv:hep-ph/9504260];
  N.~Haba,
  Prog.\ Theor.\ Phys.\  {\bf 97} (1997) 301
  [arXiv:hep-ph/9608357];
  S.~W.~Ham, J.~Kim, S.~K.~Oh and D.~Son,
  Phys.\ Rev.\  D {\bf 64} (2001) 035007
  [arXiv:hep-ph/0104144];
  S.~W.~Ham, S.~K.~Oh and D.~Son,
  Phys.\ Rev.\  D {\bf 65} (2002) 075004
  [arXiv:hep-ph/0110052];
  S.~W.~Ham, S.~H.~Kim, S.~K.~OH and D.~Son,
  Phys.\ Rev.\  D {\bf 76} (2007) 115013
  [arXiv:0708.2755 [hep-ph]].

\bibitem{Boz:2005sf}
  M.~Boz,
  Mod.\ Phys.\ Lett.\  A {\bf 21} (2006) 243
  [arXiv:hep-ph/0511072].

\bibitem{Degrassi:2009yq}
  G.~Degrassi and P.~Slavich,
  Nucl.\ Phys.\  B {\bf 825}, 119 (2010)
  [arXiv:0907.4682 [hep-ph]].


\bibitem{NMSSM:review}
For recent reviews on the NMSSM, see,
  U.~Ellwanger, C.~Hugonie and A.~M.~Teixeira,
  arXiv:0910.1785 [hep-ph];
S.~Chang, R.~Dermisek, J.~F.~Gunion and N.~Weiner,
  Ann.\ Rev.\ Nucl.\ Part.\ Sci.\  {\bf 58}, 75 (2008)
  [arXiv:0801.4554 [hep-ph]].

\bibitem{mu-problem}
J.~E.~Kim and H.~P.~Nilles,
Phys.\ Lett.\ B {\bf 138}, 150 (1984);
Y.~Nir,
Phys.\ Lett.\ B {\bf 354}, 107 (1995);
M.~Cvetic and P.~Langacker,
Phys.\ Rev.\ D {\bf 54}, 3570 (1996).


\bibitem{derm}
R.~Dermisek and J.~F.~Gunion, Phys.\ Rev.\ Lett.\  {\bf 95}, 041801 (2005);
  Phys.\ Rev.\  D {\bf 73}, 111701 (2006);
  Phys.\ Rev.\  D {\bf 76}, 095006 (2007);
S.~Chang, P.~J.~Fox and N.~Weiner,
  JHEP {\bf 0608}, 068 (2006)
  [arXiv:hep-ph/0511250].

\bibitem{LEP2003}
  R.~Barate {\it et al.}  [LEP Working Group for Higgs boson searches],
  Phys.\ Lett.\ B {\bf 565}, 61 (2003).

\bibitem{nmssm-new}
B.~A.~Dobrescu, G.~Landsberg and K.~T.~Matchev,
  Phys.\ Rev.\  D {\bf 63}, 075003 (2001);
U.~Ellwanger, J.~F.~Gunion and C.~Hugonie,
  JHEP {\bf 0507}, 041 (2005);
  V.~Barger, P.~Langacker, H.~S.~Lee and G.~Shaughnessy,
  Phys.\ Rev.\  D {\bf 73}, 115010 (2006);
S.~Chang, P.~J.~Fox and N.~Weiner,
  Phys.\ Rev.\ Lett.\  {\bf 98}, 111802 (2007);
  V.~Barger, P.~Langacker and G.~Shaughnessy,
  Phys.\ Rev.\  D {\bf 75}, 055013 (2007);
T.~Stelzer, S.~Wiesenfeldt and S.~Willenbrock,
  Phys.\ Rev.\  D {\bf 75}, 077701 (2007)
  [arXiv:hep-ph/0611242];
K.~Cheung, J.~Song and Q.~S.~Yan,
  Phys.\ Rev.\ Lett.\  {\bf 99}, 031801 (2007)
  [arXiv:hep-ph/0703149];
M.~Carena, T.~Han, G.~Y.~Huang and C.~E.~M.~Wagner,
  JHEP {\bf 0804}, 092 (2008)
  [arXiv:0712.2466 [hep-ph]];
J.~Cao, H.~E.~Logan and J.~M.~Yang,
  Phys.\ Rev.\  D {\bf 79}, 091701 (2009)
  [arXiv:0901.1437 [hep-ph]];
A.~Belyaev, J.~Pivarski, A.~Safonov, S.~Senkin and A.~Tatarinov,
  Phys.\ Rev.\  D {\bf 81}, 075021 (2010)
  [arXiv:1002.1956 [hep-ph]].

\bibitem{Sakharov:1967dj}
  A.~D.~Sakharov,
  Universe,''
  Pisma Zh.\ Eksp.\ Teor.\ Fiz.\  {\bf 5} (1967) 32
  [JETP Lett.\  {\bf 5} (1967\ SOPUA,34,392-393.1991\ UFNAA,161,61-64.1991) 24].

\bibitem{CKM}
  N.~Cabibbo,
  Phys.\ Rev.\ Lett.\  {\bf 10} (1963) 531;
  M.~Kobayashi and T.~Maskawa,
  Prog.\ Theor.\ Phys.\  {\bf 49} (1973) 652.

\bibitem{ewbg_sm_cp}
  M.~B.~Gavela, P.~Hernandez, J.~Orloff and O.~Pene,
  Mod.\ Phys.\ Lett.\  A {\bf 9} (1994) 795;~
%
  M.~B.~Gavela, P.~Hernandez, J.~Orloff, O.~Pene and C.~Quimbay,
  Nucl.\ Phys.\  B {\bf 430} (1994) 382;~
%
  P.~Huet and E.~Sather,
  Phys.\ Rev.\  D {\bf 51} (1995) 379;~
%
  T.~Konstandin, T.~Prokopec and M.~G.~Schmidt,
  Nucl.\ Phys.\  B {\bf 679} (2004) 246.

\bibitem{sm_ewpt}
  K.~Kajantie, M.~Laine, K.~Rummukainen and M.~E.~Shaposhnikov,
  Phys.\ Rev.\ Lett.\  {\bf 77}, 2887 (1996);~
%
  K.~Rummukainen, M.~Tsypin, K.~Kajantie, M.~Laine and M.~E.~Shaposhnikov,
  Nucl.\ Phys.\  B {\bf 532}, 283 (1998);~
%
  F.~Csikor, Z.~Fodor and J.~Heitger,
  Phys.\ Rev.\ Lett.\  {\bf 82}, 21 (1999);~
%
  Y.~Aoki, F.~Csikor, Z.~Fodor and A.~Ukawa,
  Phys.\ Rev.\  D {\bf 60}, 013001 (1999).

\bibitem{ewbg}
For reviews, see
A.~G.~Cohen, D.~B.~Kaplan and A.~E.~Nelson,
Ann.\ Rev.\ Nucl.\ Part.\ Sci.\  {\bf 43} (1993) 27
[arXiv:hep-ph/9302210];
%
M.~Quiros,
Helv.\ Phys.\ Acta {\bf 67} (1994) 451;
%
V.~A.~Rubakov and M.~E.~Shaposhnikov,
Usp.\ Fiz.\ Nauk {\bf 166} (1996) 493
[Phys.\ Usp.\  {\bf 39} (1996) 461]
[arXiv:hep-ph/9603208];
%
K.~Funakubo,
Prog.\ Theor.\ Phys.\  {\bf 96} (1996) 475
[arXiv:hep-ph/9608358];
%
M.~Trodden,
Rev.\ Mod.\ Phys.\  {\bf 71} (1999) 1463
[arXiv:hep-ph/9803479];
%
W.~Bernreuther,
Lect.\ Notes Phys.\  {\bf 591} (2002) 237
[arXiv:hep-ph/0205279].

\bibitem{ewbg-mssm}
  M.~S.~Carena, M.~Quiros and C.~E.~M.~Wagner,
  Phys.\ Lett.\  B {\bf 380}, 81 (1996).
  %
  D.~Delepine, J.~M.~Gerard, R.~Gonzalez Felipe and J.~Weyers,
  Phys.\ Lett.\  B {\bf 386}, 183 (1996);~
%
  P.~Huet and A.~E.~Nelson,
  Phys.\ Rev.\  D {\bf 53}, 4578 (1996);~
%
  B.~de Carlos and J.~R.~Espinosa,
  Nucl.\ Phys.\  B {\bf 503}, 24 (1997);~
%
  M.~S.~Carena, M.~Quiros, A.~Riotto, I.~Vilja and C.~E.~M.~Wagner,
  Nucl.\ Phys.\  B {\bf 503}, 387 (1997);~
  A.~Riotto,
  Nucl.\ Phys.\  B {\bf 518}, 339 (1998);~
%
  M.~S.~Carena, M.~Quiros and C.~E.~M.~Wagner,
  Nucl.\ Phys.\  B {\bf 524}, 3 (1998);~
%
  K.~Funakubo, A.~Kakuto, S.~Otsuki and F.~Toyoda,
  Prog.\ Theor.\ Phys.\  {\bf 99}, 1045 (1998);~
%
  A.~Riotto,
  Phys.\ Rev.\  D {\bf 58}, 095009 (1998);~
%
  K.~Funakubo,
  Prog.\ Theor.\ Phys.\  {\bf 101}, 415 (1999);~
%
  K.~Funakubo, S.~Otsuki and F.~Toyoda,
  Prog.\ Theor.\ Phys.\  {\bf 102}, 389 (1999);~
%
  J.~M.~Cline, M.~Joyce and K.~Kainulainen,
  JHEP {\bf 0007}, 018 (2000);~
%
  M.~S.~Carena, J.~M.~Moreno, M.~Quiros, M.~Seco and C.~E.~M.~Wagner,
  Nucl.\ Phys.\  B {\bf 599}, 158 (2001);~
%
  M.~S.~Carena, M.~Quiros, M.~Seco and C.~E.~M.~Wagner,
  Nucl.\ Phys.\  B {\bf 650}, 24 (2003);~
%
  T.~Prokopec, M.~G.~Schmidt and S.~Weinstock,
  Annals Phys.\  {\bf 314}, 208 (2004);~
  %
  T.~Prokopec, M.~G.~Schmidt and S.~Weinstock,
  Annals Phys.\  {\bf 314}, 267 (2004);~
%
  T.~Konstandin, T.~Prokopec and M.~G.~Schmidt,
  Nucl.\ Phys.\  B {\bf 716}, 373 (2005);~
%
  C.~Lee, V.~Cirigliano and M.~J.~Ramsey-Musolf,
  Phys.\ Rev.\  D {\bf 71}, 075010 (2005);~
%
  V.~Cirigliano, M.~J.~Ramsey-Musolf, S.~Tulin and C.~Lee,
  Phys.\ Rev.\  D {\bf 73}, 115009 (2006);~
%
  T.~Konstandin, T.~Prokopec, M.~G.~Schmidt and M.~Seco,
  Nucl.\ Phys.\  B {\bf 738}, 1 (2006);~
%
  V.~Cirigliano, S.~Profumo and M.~J.~Ramsey-Musolf,
  JHEP {\bf 0607}, 002 (2006);~
%
  D.~J.~H.~Chung, B.~Garbrecht, M.~J.~Ramsey-Musolf and S.~Tulin,
  Phys.\ Rev.\ Lett.\  {\bf 102}, 061301 (2009);~
 %
  K.~Funakubo, S.~Tao and F.~Toyoda,
  Prog.\ Theor.\ Phys.\  {\bf 109}, 415 (2003);~
%
  M.~Carena, G.~Nardini, M.~Quiros and C.~E.~M.~Wagner,
  Nucl.\ Phys.\  B {\bf 812}, 243 (2009);~
%
  K.~Funakubo and E.~Senaha,
  Phys.\ Rev.\  D {\bf 79}, 115024 (2009).
%

\bibitem{ewbg-2hdm}
  K.~Funakubo, A.~Kakuto and K.~Takenaga,
  Prog.\ Theor.\ Phys.\  {\bf 91}, 341 (1994);~
%
  M.~Joyce, T.~Prokopec and N.~Turok,
  Phys.\ Rev.\  D {\bf 53}, 2930 (1996);~
%
  M.~Joyce, T.~Prokopec and N.~Turok,
  Phys.\ Rev.\  D {\bf 53}, 2958 (1996);~
%
  J.~M.~Cline, K.~Kainulainen and A.~P.~Vischer,
  Phys.\ Rev.\  D {\bf 54}, 2451 (1996);~
%
  J.~M.~Cline and P.~A.~Lemieux,
  Phys.\ Rev.\  D {\bf 55}, 3873 (1997);~
%
  S.~Kanemura, Y.~Okada and E.~Senaha,
  Phys.\ Lett.\  B {\bf 606}, 361 (2005);~
%
  L.~Fromme, S.~J.~Huber and M.~Seniuch,
  JHEP {\bf 0611}, 038 (2006).

\bibitem{Funakubo:2005pu}
  K.~Funakubo, S.~Tao and F.~Toyoda,
  Prog.\ Theor.\ Phys.\  {\bf 114}, 369 (2005)
  [arXiv:hep-ph/0501052].

\bibitem{ewbg-xMSSM}
  S.~J.~Huber and M.~G.~Schmidt,
  Nucl.\ Phys.\  B {\bf 606}, 183 (2001);~
%
  A.~Menon, D.~E.~Morrissey and C.~E.~M.~Wagner,
  Phys.\ Rev.\  D {\bf 70}, 035005 (2004);~
%
  J.~Kang, P.~Langacker, T.~j.~Li and T.~Liu,
  Phys.\ Rev.\ Lett.\  {\bf 94}, 061801 (2005);~
 %
  S.~J.~Huber, T.~Konstandin, T.~Prokopec and M.~G.~Schmidt,
  Nucl.\ Phys.\  B {\bf 757}, 172 (2006);~
%
  C.~Balazs, M.~S.~Carena, A.~Freitas and C.~E.~M.~Wagner,
  JHEP {\bf 0706}, 066 (2007);~
%
  J.~Kang, P.~Langacker, T.~Li and T.~Liu,
  arXiv:0911.2939 [hep-ph];~
%
  C.~W.~Chiang and E.~Senaha,
  JHEP {\bf 1006}, 030 (2010);~
  %
  A.~Ahriche and S.~Nasri,
  arXiv:1008.3106 [hep-ph].

\bibitem{CPmixing0}
A. Pilaftsis, Phys.\  Rev.\ D {\bf 58} (1998) 096010;
  Phys.\ Lett.\ B {\bf 435} (1998) 88.

\bibitem{CPmixing1}
A.~Pilaftsis and C.E.M.  Wagner, Nucl.\ Phys.\ B~{\bf 553} (1999) 3.

\bibitem{CPmixing1.5}
D.A.  Demir, Phys.\ Rev.\ D~{\bf 60} (1999) 055006.

\bibitem{CPmixing2}
S.Y.  Choi, M.  Drees and J.S.  Lee, Phys.\ Lett.\ B~{\bf 481} (2000) 57;
M. Carena, J. Ellis, A. Pilaftsis and C.E.M. Wagner,
  Nucl. Phys. {\bf B586} (2000) 92;
M. Carena, J. Ellis, A. Pilaftsis and C.E.M. Wagner,
  Nucl. Phys. {\bf B625} (2002) 345.

\bibitem{Lee:2008eqa}
  J.~S.~Lee,
  AIP Conf.\ Proc.\  {\bf 1078} (2009) 36
  [arXiv:0808.2014 [hep-ph]].

\bibitem{Accomando:2006ga}
  E.~Accomando {\it et al.},
  arXiv:hep-ph/0608079.

\bibitem{CPH_decay}
  S.~Y.~Choi and J.~S.~Lee,
  Phys.\ Rev.\  D {\bf 61} (1999) 015003
  [arXiv:hep-ph/9907496];
  S.~Y.~Choi, K.~Hagiwara and J.~S.~Lee,
  Phys.\ Rev.\  D {\bf 64} (2001) 032004
  [arXiv:hep-ph/0103294];
  S.~Y.~Choi, M.~Drees, J.~S.~Lee and J.~Song,
  Eur.\ Phys.\ J.\  C {\bf 25} (2002) 307
  [arXiv:hep-ph/0204200].

\bibitem{cpsuperh}
  J.~S.~Lee, A.~Pilaftsis, M.~Carena, S.~Y.~Choi, M.~Drees, J.~R.~Ellis and
C.~E.~M.~Wagner,
  Comput.\ Phys.\ Commun.\  {\bf 156} (2004) 283
  [arXiv:hep-ph/0307377];
  J.~S.~Lee, M.~Carena, J.~Ellis, A.~Pilaftsis and C.~E.~M.~Wagner,
  Comput.\ Phys.\ Commun.\  {\bf 180} (2009) 312
  [arXiv:0712.2360 [hep-ph]].

\bibitem{Schael:2006cr}
  S.~Schael {\it et al.}  [ALEPH Collaboration and DELPHI Collaboration and
                  L3 Collaboration and ],
  Eur.\ Phys.\ J.\  C {\bf 47} (2006) 547
  [arXiv:hep-ex/0602042].

\bibitem{Regan:2002ta}
  B.~C.~Regan, E.~D.~Commins, C.~J.~Schmidt and D.~DeMille,
  Phys.\ Rev.\ Lett.\  {\bf 88} (2002) 071805.

\bibitem{Baker:2006ts}
  C.~A.~Baker {\it et al.},
  Phys.\ Rev.\ Lett.\  {\bf 97} (2006) 131801.

\bibitem{Romalis:2000mg}
  M.~V.~Romalis, W.~C.~Griffith and E.~N.~Fortson,
  Phys.\ Rev.\ Lett.\  {\bf 86} (2001) 2505.

\bibitem{Griffith:2009zz}
  W.~C.~Griffith, M.~D.~Swallows, T.~H.~Loftus, M.~V.~Romalis, B.~R.~Heckel and
E.~N.~Fortson,
  Phys.\ Rev.\ Lett.\  {\bf 102} (2009) 101601.

\bibitem{Ibrahim:1998je}
  T.~Ibrahim and P.~Nath,
  Phys.\ Rev.\  D {\bf 58} (1998) 111301
  [Erratum-ibid.\  D {\bf 60} (1999) 099902]
  [arXiv:hep-ph/9807501].

\bibitem{Ellis:2008zy}
  J.~R.~Ellis, J.~S.~Lee and A.~Pilaftsis,
  JHEP {\bf 0810} (2008) 049
  [arXiv:0808.1819 [hep-ph]].

\bibitem{Funakubo:2004ka}
  K.~Funakubo and S.~Tao,
  Prog.\ Theor.\ Phys.\  {\bf 113} (2005) 821
  [arXiv:hep-ph/0409294].


\bibitem{Cheung:2010ba}
  K.~Cheung, T.~J.~Hou, J.~S.~Lee and E.~Senaha,
  Phys.\ Rev.\  D {\bf 82} (2010) 075007
  [arXiv:1006.1458 [hep-ph]].

\bibitem{Casas:1994us}
  J.~A.~Casas, J.~R.~Espinosa, M.~Quiros and A.~Riotto,
  Nucl.\ Phys.\  B {\bf 436} (1995) 3
  [Erratum-ibid.\  B {\bf 439} (1995) 466]
  [arXiv:hep-ph/9407389].

\bibitem{Dicus:1989va}
  D.~A.~Dicus,
  Phys.\ Rev.\  D {\bf 41} (1990) 999.

\bibitem{Ellis:2010xm}
  J.~Ellis, J.~S.~Lee and A.~Pilaftsis,
  JHEP {\bf 1010} (2010) 049
  [arXiv:1006.3087 [hep-ph]].

\bibitem{Giudice:2005rz}
  G.~F.~Giudice and A.~Romanino,
  Phys.\ Lett.\  B {\bf 634} (2006) 307
  [arXiv:hep-ph/0510197].

\bibitem{Li:2008kz}
  Y.~Li, S.~Profumo and M.~Ramsey-Musolf,
  Phys.\ Rev.\  D {\bf 78} (2008) 075009
  [arXiv:0806.2693 [hep-ph]].

\bibitem{KL} I.B. Khriplovich and S.K. Lamoreaux, {\em CP Violation
  Without Strangeness} (Springer, New York, 1997).

\bibitem{Pospelov:2005pr}
  M.~Pospelov and A.~Ritz,
  Annals Phys.\  {\bf 318} (2005) 119.

\bibitem{Ibrahim:1997gj}
  T.~Ibrahim and P.~Nath,
  Phys.\ Rev.\  D {\bf 57} (1998) 478
  [Erratum-ibid.\  D {\bf 58} (1998\ ERRAT,D60,079903.1999\ ERRAT,D60,119901.1999) 019901]

\bibitem{Ellis:1996dg}
  J.~R.~Ellis and R.~A.~Flores,
  Phys.\ Lett.\  B {\bf 377} (1996) 83.

\bibitem{qcdsumrule1}
  M.~Pospelov and A.~Ritz,
  Phys.\ Rev.\ Lett.\  {\bf 83} (1999) 2526;\\
  M.~Pospelov and A.~Ritz,
  Nucl.\ Phys.\  B {\bf 573} (2000) 177.

\bibitem{qcdsumrule2}
  M.~Pospelov and A.~Ritz,
  Phys.\ Rev.\  D {\bf 63} (2001) 073015.

\bibitem{Demir:2002gg}
  D.~A.~Demir, M.~Pospelov and A.~Ritz,
  Phys.\ Rev.\  D {\bf 67} (2003) 015007.

\bibitem{Demir:2003js}
  D.~A.~Demir, O.~Lebedev, K.~A.~Olive, M.~Pospelov and A.~Ritz,
  Nucl.\ Phys.\  B {\bf 680} (2004) 339.

\bibitem{Olive:2005ru}
  K.~A.~Olive, M.~Pospelov, A.~Ritz and Y.~Santoso,
  Phys.\ Rev.\  D {\bf 72} (2005) 075001.

\bibitem{Ellis:2011hp}
  J.~Ellis, J.~S.~Lee and A.~Pilaftsis,
  JHEP {\bf 1102} (2011) 045
  [arXiv:1101.3529 [hep-ph]],
and references there in.


\bibitem{Lebedev:2004va}
  O.~Lebedev, K.~A.~Olive, M.~Pospelov and A.~Ritz,
  Phys.\ Rev.\  D {\bf 70} (2004) 016003.

\bibitem{Engel:2003rz}
  J.~Engel, M.~Bender, J.~Dobaczewski, J.~H.~De Jesus and P.~Olbratowski,
  ``Time-Reversal Violating Schiff Moment of 225Ra,''
  Phys.\ Rev.\  C {\bf 68}, 025501 (2003).

\bibitem{Hebbeker:1999pi}
  T.~Hebbeker,
  Phys.\ Lett.\  B {\bf 470} (1999) 259
  [arXiv:hep-ph/9910326].

\bibitem{Harnik:2003rs}
R.~Harnik, G.~D.~Kribs, D.~T.~Larson and H.~Murayama,
Phys.\ Rev.\  D {\bf 70}, 015002 (2004)
[arXiv:hep-ph/0311349].

\bibitem{Semertzidis:2003iq}
  Y.~K.~Semertzidis {\it et al.}  [EDM Collaboration],
  AIP Conf.\ Proc.\  {\bf 698} (2004) 200.

\bibitem{Willmann}
L. Willmann, K. Jungmann, H. W. Wilschut, ``Searches for permanent
electric dipole moments in Radium Isotopes'',
Letter of Intent to the ISOLDE and Neutron Time-of-Flight Experiments Committee
for experiments with HIE-ISOLDE, CERN-INTC-2010-049 / INTC-I-115.


\bibitem{Li:2008ez}
  Y.~Li, S.~Profumo and M.~Ramsey-Musolf,
  Phys.\ Lett.\  B {\bf 673} (2009) 95
  [arXiv:0811.1987 [hep-ph]].

\bibitem{Chang:1998uc}
  D.~Chang, W.~-Y.~Keung, A.~Pilaftsis,
  Phys.\ Rev.\ Lett.\  {\bf 82 } (1999)  900
  [hep-ph/9811202]
[Erratum-ibid.\  {\bf 83} (1999) 3972].

\bibitem{More_twoloops}
  A.~Pilaftsis,
  Phys.\ Lett.\  {\bf B471 } (1999)  174
  [hep-ph/9909485];
  D.~Chang, W.~-F.~Chang, W.~-Y.~Keung,
  Phys.\ Lett.\  {\bf B478 } (2000)  239
  [hep-ph/9910465].


\bibitem{future1}
K. Cheung, T.J. Hou, J.S. Lee, and E. Senaha, in preparation.

\end{thebibliography}
\end{document}